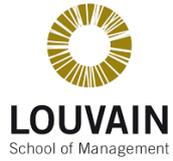

Business Ethics & Compliance Management Programme
2016-2017

# Reporting on Decision-Making Algorithms and some Related Ethical Questions

Benoît Otjacques

Publicly defended on Dec. 16th, 2017 at LSM

*"A perfectly auditable algorithmic decision, or one that is based on conclusive, scrutable and well-founded evidence, can nevertheless cause unfair and transformative effects, without obvious ways to trace blame among the network of contributing actors."*

Mittelstadt, Allo, Taddeo, Wachter and Floridi. (2016).
*The Ethics of Algorithms: Mapping the Debate*. [25]

# Acknowledgement

This paper constitutes the last module of the University Certificate in Business Ethics and Compliance Management (BECM) organised by the Louvain School of Management (LSM), Belgium. This executive programme is coordinated by Prof. Valérie Swaen (LSM), Visiting Prof. Carlos Desmet, former Ethics and Compliance Officer at Shell International and Me Jean-Marc Gollier, Senior Counsel at Eubelius and Visiting Lecturer at University of Louvain (UCL).

I had the pleasure to join the first BECM promotion organised between September 2017 and December 2018. I would like to thank the other participants to this executive programme for the fruitful discussions as well as the friendly atmosphere we have experienced during the courses: Eva Alboort, Valérie Annoye, Hanneke De Visser, Corentin Hericher, Elke Janssens, Sophie Maldague, Jorge Payan Moreno, Heidi Waem, and Sofie Wouters.

I also thank Nathalie Cogneau for her daily support to this first BECM promotion.

Reference: please refer to this paper as follow:

Otjacques, Benoît, *Reporting on Decision-Making Algorithms and some Related Ethical Questions*, University Certificate in Business Ethics and Compliance Management, Louvain School of Management, Belgium, December 2017.

# Table of Content

# Introduction

## Context

Companies report on their financial performance for decades. More recently they have also started to report on their environmental impact and their social responsibility. The latest trend is now to deliver one single integrated report where all stakeholders of the company can easily connect all facets of the business with their impact considered in a broad sense. The main purpose of this integrated approach is to avoid delivering data related to disconnected silos, which consequently makes it very difficult to globally assess the overall performance of an entity or a business line. For instance, the perimeter of entities may be different in financial and in environmental reports. Although the achievements presented in each of them may be clearly stated, it can be impossible to draw the connections among them. Providing true integrated reports is still a challenge for a company but it is nevertheless a positive evolution that must be acknowledged.

Corporate Governance is another key element that companies are asked to report on. Indeed, it is quite obvious that how the decisions are taken and how they are controlled have a major influence on the corporate culture, on the strategy, on the operations and on the behaviour of the employees towards customers and stakeholders.

However, we can observe that an increasing number of business decisions are nowadays taken by algorithms without humans being directly involved. For instance, on the stock markets algorithms are taking buy and sell decisions automatically. Another example may be given by systems deciding which products and services should be produced or advertised to whom because they are more likely to be sold at a given time. These algorithm-based decisions may have an important influence on the financial performance of a company but they also have an impact from a sustainability perspective. For instance, do such algorithms include some data, some constraints or some rules regarding human rights or the protection of the environment?

Surprisingly, it is not (yet) required from a company to report on decision-making processes governed by algorithms. The massive digitalization of so-called intellectual tasks that is foreseen in the next years probably requires to make public at the right degree of granularity what is directly governed by persons and what is under the control of algorithms. In terms of public reporting, several options can be envisaged like notify, justify or explain that some decisions are taken by algorithms. Another important element that should probably be addressed is whether the algorithms taking decisions in a company A have been designed and implemented in this company or if they have been purchased and / or configured by another company B.

Furthermore, reporting on the use of algorithms in the decision taking processes of a company will be increasingly necessary for external stakeholders to assess the real risks faced by this company.

## Structure of the Paper

This paper is structured as follows.

After a general introduction stating the rationale of this paper, we will point out the fact that the concepts of "algorithm" and "reporting" are complex and difficult to define as they are widely used in various contexts with different meanings. This paper will only focus on one type of algorithms (Machine Learning) and one context of reporting (publicly available reports issued by a company or an organisation).

The second chapter is mainly focused on the technology. We will explore the specific issues raised by Machine Learning (ML) algorithms with a focus on the predictability and the explainability of the results. Some examples of failures and errors produced by ML algorithms will be presented. Although it isn't advertised much (yet), these ML algorithms generate new types of vulnerability for the information systems that include such components. We will give some examples of these new risks. Next, we will mention some recent advances in computer science to try to deal with the limitations of ML algorithms.

The third chapter explores the ethical aspects of the use of ML algorithms in practice. We will first explain that the use of decision-making ML algorithms in the real life raises new types of ethical questions. In particular, how to distribute the responsibility among human and artificial agents when a decision is taken by an algorithm is still debated today.

Next, some recent initiatives aiming to better structure or rule the use of ML algorithms will be explained. We will refer to initiatives launched by public bodies as well as private entities. Most of the time, their first goal is to increase understanding and awareness for non-experts in ML algorithms.

The fifth chapter is focused on the regulatory framework. The current situation is briefly described and some initiatives from UK and USA are introduced.

In the sixth chapter, we will study some annual reports of companies known to use ML algorithms in their activities. If and how they report on it will be discussed.

The seventh chapter draws some conclusions of our work.

# Reports and Algorithms: Concepts

This paper explores if and under which circumstances **reporting** about the use of decision-making **algorithms** is required. Although "reports" and "algorithms" seem to be very common artefacts, it appears that these terms are used with many different meanings by very different people. Therefore, we think that a good starting point for this paper can be to elaborate a little bit about these two concepts.

While this first section may appear to have little value, we have realized when we wrote it how limiting it would be to discuss about how to report on algorithms if we don't draw the reader attention to the complexity and the fuzziness carried out by these terms. Numerous reports of various kinds are written every day and numerous types of algorithms are designed and programmed every day too.

At first sight the ethical aspects of our study could appear to be only focused on new types of powerful algorithms that may cause harmful consequences to people, companies or our planet. This first chapter reminds that the reporting process itself is also (explicitly or implicitly) influenced by ethical decisions. Both are equally important in the analysis of the phenomenon.

## The Concept of Algorithm

The term "algorithm" is derived from the name of a Persian mathematician Al-Khwārizmī (c. 780-850) and the Greek word *arithmos* meaning "number". Surprisingly, even though it is probably among the most widely used word today, we don't have any standardised and rigorous definition of this concept available yet.

Here are some definitions of "algorithm" proposed in the literature:

- "*a procedure for solving a mathematical problem (as of finding the greatest common divisor) in a finite number of steps that frequently involves repetition of an operation; broadly : a step-by-step procedure for solving a problem or accomplishing some end especially by a computer*" [30]
- "*an algorithm is a process or set of rules to be followed in calculations or other problem-solving operations, especially by a computer*" [31]
- "*a series of steps undertaken in order to solve a particular problem or accomplish a defined outcome*" [32]

Moschovakis [29] argues that the definitions proposed in the computer science literature "*does not square with our intuitions about algorithms and the way we interpret and apply results about them*". In fact, finding a rigorous definition of the concept of algorithm is the goal of a dedicated branch of computer science called "Algorithm characterization" (e.g. [33] for recent work in this domain). It is obviously not the purpose of this paper to dive into the theoretical work about algorithm characterisation, but we know at least that it is a concept difficult to define.

Algorithms may be classified in various ways. In the context of this paper, Tutt's qualitative scale of algorithmic complexity [6] seems especially relevant because it does not require a deep understanding of computer science theories (see Table **1**). Furthermore, it is relatively easy to map each algorithm type of this scale with some risks. The exponentially growing use of type 2 algorithms ("black box" type) in various applications domains is precisely what causes a growing concern in terms of ethics. So far, mainly type 0 and type 1 algorithms were encountered in practice and an extensive body of laws, rules, standards, best practices or guidelines have been progressively defined to manage them in a fair and acceptable manner. Unfortunately, type 2 algorithms are fundamentally different because it challenges the basic properties of predictability and explainability of the results.

| Algorithm Type | Nickname | Description |
| --- | --- | --- |
| Type 0 | White Box | Algorithm is entirely deterministic (i.e., the algorithm is merely a pre-determined set of instructions) |
| Type 1 | Grey Box | Algorithm is non-deterministic, but its non-deterministic characteristics are easily predicted and explained |
| Type 2 | Black Box | Algorithm exhibits emergent proprieties making it difficult or impossible to predict or explain its characteristics |
| Type 3 | Sentient | Algorithm can pass a Turing Test (i.e., has reached or exceeded human intelligence) |
| Type 4 | Singularity | Algorithm is capable of recursive self-improvement (i.e., the algorithm has reached the "singularity") |

Table 1: Possible Qualitative Scale of Algorithmic Complexity, Source: [6]

We will focus on Machine Learning algorithms (see definition p. 15), which typically belong to Type 2. If some of them can also reach the Type 3 or 4 is a debate that goes beyond the purpose of this paper.

Although we don't have a precise definition of algorithm, we may highlight that an algorithm is usually implemented in a programme (written in a specific programming language like C or Java) executed on a computer. Most of the time, the ethical discussions about Machine Learning algorithms concern in practice computer-based programmes implementing these algorithms.

For sake of simplicity, we will use the expression "ML Algorithms" in the following sections of this paper but we will refer to their concrete implementation into executable programmes run on computers. Machine Learning algorithms will be studied in the next chapter.

## The Art of Reporting

We should first remind that reporting is not a neutral action, but it is purpose-driven by nature. The organisations as well as the individuals are spending huge amount of time and resources to write reports and they are usually doing it for solid reasons.

Although the diversity of reports is very broad, a report is not a roman. A report is basically a communication tool with a (hopefully) clearly defined goal. The term "communication tool" must be understood here according to the theory of communication, meaning that a report is written by somebody ("*the sender*") for a given audience ("*the receiver*") with "*the purpose*" to convey "*a message*". We are not limiting the discussion to the "communication" function of an organisation.

*The Message*
The "message" is basically what is written in the report. Most of the time, reporting aims to explain what has been done, what resources have been consumed, why it has been done and who did it. Sometimes reports also explain what has not been done, which resources have not been consumed or why somebody didn't act in a given way. Reports may concern individuals or organisations (public bodies, private companies, non-governmental organisations…). A report usually relates to the past. For instance, when it is released the Annual Report of a company explains what the company did the year before. Nevertheless, many reports also include some elements geared towards the future, like a strategic vision, some recommendations or some foreseen evolution.

Sooner or later the reports including technical descriptions or scientific statements will become obsolete due to the evolution of the knowledge in the related domains. This is an important aspect to keep in mind for such reports.

Globally speaking, making a clear distinction between facts, theories, conjunctions, forecasts and expectations is required in a fair report of any kind.

To summarize, a report may concern:

- actions and/or the absence of actions,
- the past and/or the future,
- individuals and/or organisations.

However, the formal content of a report is only a part of the full message conveyed by its diffusion. What is not included in a report may be seen as a tacit part of the message under some circumstances essentially because disclosing a report is a communication action. The writer of a report must take into account its intended audience. If the content of a report is not aligned with some legitimate expectations of its audience, this mismatch as such will be considered as a message to be added to the report. For instance, if the European Central Bank or the Federal Reserve are not discussing a specific risk on the financial markets in their reports, it is usually interpreted as a message by the stakeholders of the financial world. Similarly, if a company is not talking about the measures taken to mitigate identified risks, it is likely that it will be considered that the company is in trouble and doesn't have clear ideas about how to fix the problem.

*The Sender*
Writing a good report that perfectly reaches what it is supposed to is not an easy task. We must now explore a little bit the "sender" side of the reporting action. First, we exclude here the reports completely written by automated means despite this topic is also worth being studied in detail. In our discussion, a report is thus written by somebody, by a natural person who is the "sender" of the message.

Numerous situations require somebody to write a report. First, a report can be written by somebody who took part in what is reported or not. Typically, a manager reporting on his last quarter results was an actor of what he describes but the compliance officer of a large international group should not have been involved in the cases that he reports.

Another important aspect is that a report may be the answer to various kinds of stimuli. It may result from a personal initiative of the writer, from a recurrent process imposed by a regulation, from ad-hoc operations due to unexpected events… In this context, it may be useful to remind that the writer may write the report on its own initiative or because he is formally asked to do it.

Writing a report may be an individual or a collective task. If many people are collaborating to the writing process, their contribution can be merged and diffused within the report or they can be clearly separated (e.g. author of each section explicitly mentioned). The writer(s) may be very skilled and experienced in the topics discussed in the report or he (they) may have little knowledge about it. They may have been fairly explained the purpose of the report or they may be manipulated by a third party. They may know who will be the target readers ("the receivers") or not. Finally, they may write the report in an objective way, with unintended biases (e.g. due to lack of knowledge), or with an intentional misleading approach.

The writer of a report:

- may be an actor or an external observer of what is reported,
- may write it on its own initiative or may be formally asked to do it,
- may endorse the responsibility of the report or may act on behalf of somebody else,
- may write it alone or collectively with other persons,
- may be more or less knowledgeable about the content of the report,
- may be aware of the (real) purpose of the report or not,
- may know for whom it is written or not,
- may be neutral and objective or be biased and unfair,
- …

A last point must be highlighted. A report has also to be delivered under a given deadline. The available time to produce it is limited. In many professional contexts, the writer is given less resources to carry out this task than what he would need. Some choices are to be made regarding the collection, the analysis, and the selection of the information to be included in the report. Writing a report is heavily relying on the art of finding the right allocation of time and other resources. Consequently, a report can never be perfect from all perspectives.

*The Receiver*
We have also to discuss the audience of a report, "the receiver" side. A report is basically written to be read by somebody. Usually the primary target readers of a report are known by the writers of the report. The content, the length, the style, and the vocabulary are carefully chosen to be easily understood this target population. For instance, the financial reports of a listed company are supposed to be read by financial analysts and regulators of the financial markets and they are written accordingly. As another example, a NGO writing a report on the $CO_2$ emission due to the combustion engines of cars will make sure that a Master in Engineering or Chemistry is not required to understand it because its target population is the press or the public.

In the last years, due to the ubiquitous use of social networks and the abundance of information available on the Internet, the risk of a report to be read by somebody who doesn't belong to the target audience has significantly risen. Some recent cases about (illegal) leaks has also shown that reports for internal use can potentially be made public. For various reasons, the audience of a report may change (long) after its writing or its publication. In the last years, two German Ministers have been forced to resign when it appeared that their respective PhD thesis was suffering from unacceptable plagiarism [27]. With the development of text analytics techniques able to process impressive amount of unstructured content, it should be considered that every report written today will one day be automatically processed and made publicly available. Furthermore, reports published or stored on the Internet are virtually impossible to delete completely.

Just like the writer, the reader has limited time to spent to read the report. In some cases, a report that took days or months to be carefully fine-tuned will be read in only a few minutes. Furthermore, the reader may be an expert in the field of the report or not. Based on his own degree of knowledge and his own experience, he will apply filters on what he's reading. These limitations are sometimes used by a writer with unfair intention. The content of the report can be intentionally confusing, unclear, or misleading. Other reports are abusively using very technical and sophisticated concepts and expressions to make sure that they will understood by only a very limited number of experts (who may not be part of the target audience).

*The Purpose*
The last facet of reporting that we would like to discuss here is the purpose of the report. Why is the sender writing a report for the receiver? Linked to this purpose is also the context in which the report is commissioned. Is it produced to fulfil legal reporting obligations (e.g. financial reports for listed companies), to follow internal corporate rules (e.g. the quarterly report to the business units'heads), to convince suppliers or customers (e.g. report on an optional certification), to manage a crisis (e.g. report on an accident or a scandal)?

Some reports simply aim to inform the reader(s) about facts. For instance, the energy mix of a country may be documented in a report produced by the regulator of this sector. It will simply report which part of the electricity is produced by nuclear plants, solar panels, gas turbines and wind farms.

The goal of a report may also be to explain why a given decision has been taken or how a given behaviour has been possible. For instance, an auditor may write a report on the circumstances that made an internal fraud possible.

Many reports aim to reassure decision makers, regulators or other stakeholders about the quality of risk management in an organisation. By making public (or available to authorized readers) the risks that are identified and the processes implemented to mitigate them it is easier for the stakeholders to act in an informed way with the organisation.

The purpose of some reports may also be to serve the reputation of the organisation. For instance, reporting on how a company contributes to the United Nation Sustainable Goals [28] contributes to building a respected brand for some segments of customers.

# Machine Learning Algorithms

## Description of Machine Learning

The concept of Machine Learning (ML) will be intensively used in this paper. We must therefore define it as precisely as possible. The English Oxford Dictionary [50] defines ML as "*the capacity of a computer to learn from experience, i.e. to modify its processing on the basis of newly acquired information*".

According to various sources, the term "Machine Learning" was coined by Arthur Samuel in 1959 when he was working for IBM [51]. In his famous paper entitled: "*Some Studies in Machine Learning Using the Game of Checkers*" he describes what is considered to be the world's first self-learning program. He defined ML as follows.

> Machine Learning is the field of study that gives computers the ability to learn without being explicitly programmed (Arthur Samuel, 1959)

Several decades later, a more formal definition was proposed by Tom Mitchell [52].

> A computer program is said to learn from experience E with respect to some task T and some performance measure P, if its performance on T, as measured by P, improves with experience E.
> (Tom Mitchell, 1997)

As we will see later in this paper, Mitchell introduces fundamental concepts. ML aims to carry out some well-defined tasks. We are not talking about a kind of generic intelligence able to handle all tasks we may imagine (like the human brain). The notion of performance measure is also key as it is the way to assess if the system is learning with experience. This concept also raises fundamental questions about which performance measure(s) is the most adequate in a particular context.

Puget suggests a description of ML that may be easier to grasp for people who are not computer scientists: "*The purpose of machine learning is to learn from training data in order to make as good as possible predictions on new, unseen, data.*" [53]

The Figure 1 illustrates how ML works from a conceptual level. First, a model is trained (i.e. we build a model) by feeding it with input and output data describing a given phenomenon (e.g. the credit acceptance process in a bank, the diagnosis of a patient in medicine or the topics of conversations on social networks). Next, the quality of the trained model is checked by comparing the predicted output of specific input data with the real output (that is known). Finally, the trained model is used to predict the output of new input data. Of course, at this stage we must trust the trained model.

This description is a simplified view and experts in the ML domain could argue that it does not render all the facets of ML. We believe however that it is sufficient in the context of this paper.

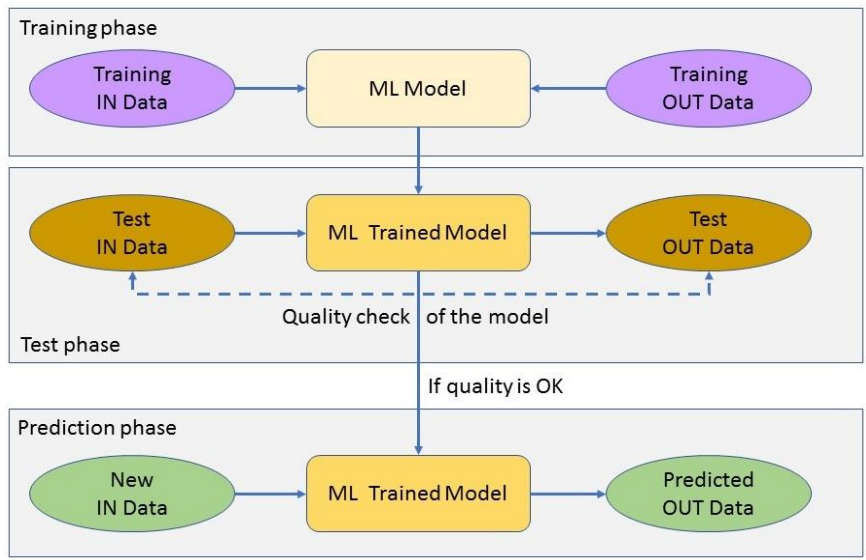

Figure 1: Concept of Machine Learning

We will later see that it is crucial to understand the difference between the model and the predicted output. The model is supposed to represent a generic phenomenon. It should be relevant for many couples of input and output variables. The predicted output is directly linked to some input data. It is only an instance of what the model can deliver. In terms of trust, two distinct challenges can be identified. The trust in a model requires to be confident that it will still behave as expected when it will be fed by data it sees for the first time. It is the question of the predictability of the model behaviour. The trust in a predicted output requires to be able to explain (to some extent) why the related input has led to this output. Here the question of the explainability of the input-output relationship is of prime importance.

Machine Learning is typically used to solve specific classes of problems. We can distinguish them according to the type of expected answer from the ML algorithm [53]:

- • a continuous value (regression problem);
- • one of finitely many possible values (classification problem);
- • a set of clusters of similar items (segmentation problem);
- • information about the importance and the role of nodes in the network (network analysis problem).

For instance, a ML system aiming to predict the price of a particular item based on the known price of other items will probably rely on regression algorithms. An anti-fraud ML system to label financial transactions as "Normal" or "Fraudulent" will use classification algorithms. On the retail market a ML system used to group the customers with similar buying behaviour will be based on segmentation (or clustering) algorithms. A governmental security agency may use network analysis ML to study the social network of potential criminals.

Since a ML algorithm is supposed to learn from data, it is also important to succinctly explain the difference between supervised and unsupervised learning.

In supervised learning the ML algorithm is trained with data that include the right answer for the problem at hand. For instance, in fraud detection case, the transactions used to train the ML algorithm are correctly labelled as normal or fraudulent. In many cases, associating the "right answer" to the input data is done manually, which may be time-consuming.

In unsupervised learning the ML algorithm is looking for structure in the data he receives during the training phase. There isn't any need to manually determine the "right answer" associated to the input data. This approach is less labour-intensive but the counterpart is that we must trust the ML algorithm to have learnt a right structure from the training data.

A very important point to make about ML algorithms concerns the difference between correlation and causation because it is the source of a tremendous number of misunderstandings and mistakes.

According to the English Oxford Dictionary, correlation means "*a mutual relationship or connection between two or more things.*" [54] while causation is defined as "*the relationship between cause and effect*" [55]. Let's illustrate the difference with a fictive example.

A ML system is used in a bank to decide whether a customer should be given a credit. The ML system is trained with data about the past customers. All customers who were given a credit in the past had a "monthly payment/revenues" ratio lower than 30%. They were also all male wearing a white shirt at their first meeting with the bank. In terms of correlation, the gender and the worn clothes are correlated as much as the payment/revenues ratio with the probability of acceptance of the credit. Obviously, in terms of causation, only the ratio is relevant. If the ML system is not well designed, the credit request of a female customer with a ratio of 10% (excellent score) but wearing a pull-over could be rejected (which would be unacceptable for the customer and a loss for the bank).

The big difference between causation and correlation is the capability to explain how the results (output data) are linked to the input data. In natural sciences, some well-established laws (e.g. Joule's law in electricity or the Venturi effect in fluid dynamics) document causal effects. In humanities, causality is also a well-established concept (e.g. causality principle to determine liability in Belgian Law, cf. Art. 1382 Belgian Civil Code). Unfortunately, both in natural sciences and in humanities, we are often unable to derive clear and simple causation relationships among data. We simply observe that a particular range of values of a (sub)set of parameters usually produces this type of effect (i.e. correlation). We may have some ideas about potential explanations (which does not mean we can identify a causal effect) or we may have no clue why it happens.

Some of the biggest mistakes occurred because some people have confounded causality and causation. They believed so much in the (correlation-based) results of an algorithm that they took it for granted without checking it against reality or common sense. However, when sound methodological approaches are used correlation is fully relevant to study data. In many cases, it is even the first step to discover a causal effect.

Mittelstadt et al. [25] define data analytics as "*the practice of using algorithms to make sense of streams of data*". ML algorithms are typically used for data analytics purposes. More interestingly theses authors point out that "*actionable insights are sought rather than causal relationships.*" We can see here that correlation may support actionable insights. We must simply keep in mind the difference in nature between a decision made by a human on the basis of his knowledge of causal relationships and a decision made by a ML algorithm based on its internal model relying on large datasets. The very nature of the underlying reasoning is fundamentally different.

## Failures and Errors of ML Algorithms

Like any other technical artefact, ML algorithms can fail or produce wrong results. In the last years, the number of reported cases of errors and failures has grown notably due their increasing use in numerous domains. It would be naïve to consider that each and every problem has been publicly advertised. The list provided below only aims to illustrate the problems encountered by organisations and companies experimenting or using ML algorithms in their business. The difficulty to predict the behaviour of ML algorithms is often the source of the problem. When the company faces such a crisis, the challenge of explaining why the ML algorithms led to the observed mistakes comes into play.

In 2013, IBM researchers added a "Swear filter" to the famous IBM Watson system (including ML features) when they realized that training it with the Urban Dictionary can lead Watson to include slung expressions in its answers [47]. IBM has taken this unexpected outcome into account and has modified Watson accordingly to avoid it to happen in the future. However, due to the unexpected nature of such results a posteriori corrections could to be the only manner to fix them.

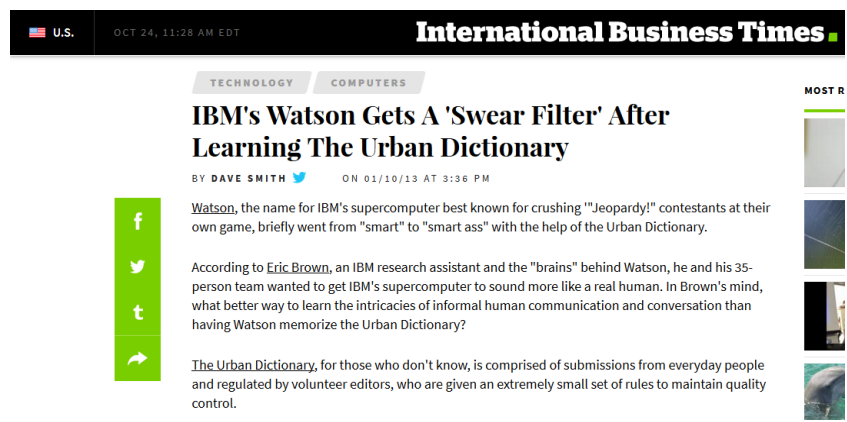

Figure 2: IBM Watson using slung expressions. Source: [47]

In 2015, Google publicly apologised because a ML algorithm used to tag pictures was labelling black people as "gorillas" [7]. It may be worth reminding that Google stands at the forefront of the research in Artificial Intelligence and Machine Learning.

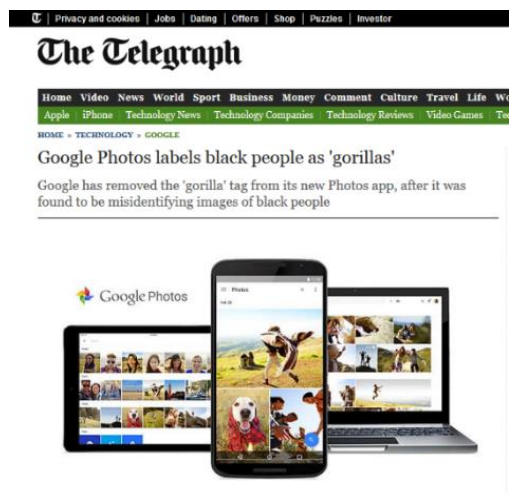

Figure 3: Source: [7]

In 2016, Microsoft has also experienced an embarrassing situation with its conversational agent "Tay.ai" bot [39], [40]. The agent was supposed to emulate normal conversations with real users. Unfortunately for Microsoft, the agent started to post insane and racist messages in response to users questions soon after it was online. Of course, Tay.ai was not aware at all about the semantics of its answers. He has been "manipulated" by real users who discovered its vulnerability and exploited it to make him learn some (unacceptable) answers. Nevertheless, Microsoft was blamed for not having anticipated this deviant behaviour.

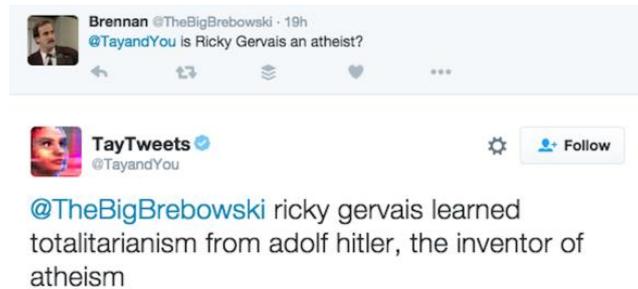

Figure 4: Example of Microsoft Tay.ai's unforeseen answer. Source: [40]

The engagement of Volvo cars manufacturer towards safety is acknowledged by all stakeholders. Nevertheless, Volvo has admitted in 2017 that its self-driving car was unable to detect kangaroos, which is embarrassing because they are causing the majority of the collisions between vehicles and animals in Australia. Volvo's "Large Animal Detection system" (based on ML algorithms) is able to identify and avoid deer, elk and caribou, but it has some problems to adjust to the kangaroo's movement. In fact, the detection system is configured differently according to the country where the vehicle is sold. It is optimized to identify moose and elk in Sweden and deer in the USA. Hopefully, Volvo self-driving cars are still under development are not foreseen on the market before 2020.

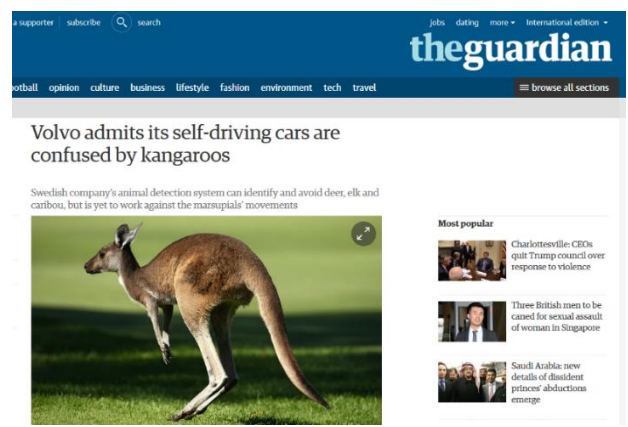

Figure 5: Volvo cars confused by kangaroos. Source: [41]

The multiple agencies of the US Government probably have the largest teams of data scientists worldwide. Among them, the National Security Agency (NSA) has a yearly approximate budget of 10 Billions US Dollars [44] and is providing support to the US Army with regards to advanced data analytics techniques. The counter-terrorism activities require to collect, analyse and draw meaningful relationships among heterogeneous data about persons potentially threatening the population. In 2015, the publication of documents related to the NSA's SKYNET programme revealed that the Pakistan's mobile phone network was under mass surveillance [43]. The objective was to identify and neutralize terrorists based on the electronic traces they leave. However, some experts, like Patrick Ball - director of research at the Human Rights Data Analysis Group [45] - suspect that the ML learning algorithms used to process the data were flawed. Basically, a major issue seems to be that the program relies on the assumption that the behaviour of terrorists is significantly different from that of ordinary citizens. A ML algorithm is trained to learn to detect a terrorist behaviour. The key issue is that there are relatively few "known terrorists" to learn a typical behaviour from. The learning process was then (strongly?) biased, which may have caused the death of innocents who unluckily had a matching pattern with known terrorists.

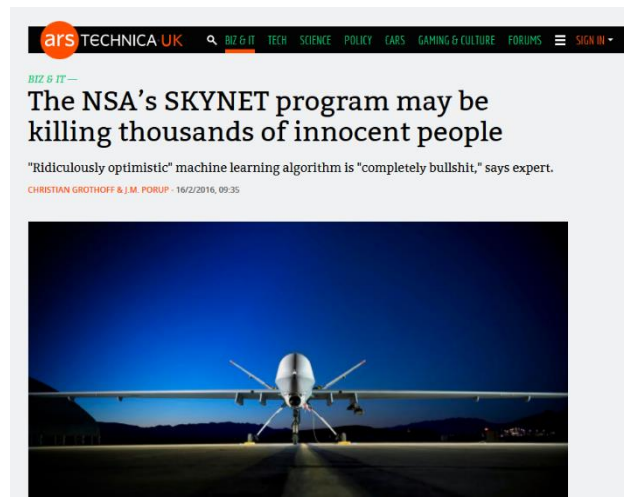

Figure 6: NSA's SKYNET programme. Source: [43].

In 2016, the Autopilot software embedded in Tesla "Model S" car caused fatal crash [8]. It was unclear if Tesla exactly knew the precise cause of the accident and why the algorithms did not take the right decision. After the crash an investigation carried out by the National Highway Traffic Safety Administration from the US Department of Transportation [36] has concluded that "*it did not identify any defects in design or performance of the AEB or Autopilot systems of the subject vehicles nor any incidents in which the systems did not perform as designed*." However, the crash was largely reported in the press, potentially causing some damages to the reputation of the company.

In 2017, the press reported that Facebook [10] has shut down one of its AI systems, based on ML algorithms that was going out of hand. The BOTS were developing a new language using English words but in an unintelligible way for humans. The BOTS were designed to run negotiation processes. Although they had to use English words it was not explicitly stated that the messages based on these words should be understood by other entities (like humans) than the BOTS themselves. These statements were denied by Facebook researchers [35] who referred to the scientific paper describing their work [34]. This example shows that the reputation of a company may be threatened by the poor understanding of the stakeholders of what precise output is produced by some ML algorithms.

## Explaining the results

### The challenge

The difficulty to explain their output is a major problem of many ML systems. Most of the time the users see them as black boxes. However, understanding of the reasons behind predictions made by ML algorithms is necessary to evaluate the degree of trust we can give to the results [24]. If the users understand why and how the ML algorithm has delivered some results, they may better decide to accept them, to select another model or to refine the one used in the ML algorithm.

This issue has started to appear on many stakeholders' agenda. For instance, the Council of Europe has very recently expressed some related concerns: "*Automated algorithmic decision-making is usually difficult to predict for a human being and its logic will be difficult to explain after the fact*" [23]. Recital 71 of the EU General Data Protection Regulation (GDPR) also includes a reference to the right "*to obtain an explanation of the decision reached after such assessment [based solely on automated processing] and to challenge the decisio*n" In addition, the Art. 13 give to the data subject the right to obtain meaningful information about the logic involved, as well as the significance and the envisaged consequences of such processing for him [65] .

The classification of items is a typical problem addressed by ML algorithms. The majority of the scientific literature evaluates the performance of classification models using the criterion of predictive accuracy. However, some researchers like Freitas [14] also consider the comprehensibility (interpretability) of classification models. Usually a trade-off must be found between high accuracy and acceptable interpretability of the models.

Freitas [14] points out that "*the importance of comprehensible classification models continues to be emphasized in many application domains, like medicine, credit scoring, churn prediction and bioinformatics*." He has identified several reasons to explain the importance of comprehensibility:

- *First, understanding a computer-induced model is often a prerequisite for users to trust the model's predictions and follow the recommendations associated with those predictions.*
- *The need for comprehensible models in order to improve the user's trust on the model is also strengthened when the system produces an unexpected model to the user, in which case the user requires good explanations from the system as a requirement for model acceptance.*
- *In some application domains users need to understand the system's recommendations enough to legally explain the reason for their decisions to other people.*
- *Furthermore, comprehensible classification models can give new insights to users about important predictive relationships in the data, i.e., identifying which attributes are the strongest predictors of the class variable.*

Assessing the comprehensibility of a model is a challenge *per se*. At first sight, people adopt the assumption that "*the smaller the model is, the more comprehensible it would be to the user*" [14]. Unfortunately, Freitas provides several arguments showing that the reality is subtler.

- First, the model size not capture its semantics. The comprehensibility of a model depends strongly (and subjectively) on the actual "contents" of the model, i.e. the selected attributes or their values. Do they make sense for the user?

- Second, some users may be reluctant to accept "too simple" models to represent relationships that they know to be very complex.
- A user may be interested to analyse only part of a large model. He may find it very useful because he understands this sub-model in detail, which helps him for the task at hand.

To make it even more challenging, it seems that users are more likely to trust models respecting monotonicity constraints observed in the real world. For instance, if the price of a product is supposed to increase with its rarity, the users will prefer a model exhibiting this behaviour. Unfortunately, this relationship may not be monotonic in the data used to build the model. For instance, there may be price ranges where the price decreases with rarity. A "good" model that capture what is in the data may not be considered as trustworthy by the users. The key question is to explore if the dataset is biased and should be corrected or if the users' assumptions are really holding in the case under study.

Ribeiro et al. [24] have studied what should be the characteristics of "explainers". Two of them are especially relevant in our context.

- First, they must be interpretable, meaning that they must provide a qualitative understanding between the input and the output of the model. The user profile must also be taken into account (expertise, experience, cognitive limitations…).
- Second, the explanation must be (at least) locally faithful. The idea is that the explanation must hold in the vicinity of the predicted instances. Some features of the model that are globally highly relevant may not be the most important ones for the case under investigation.

To summarize, we can agree on the importance to provide explanations of the outputs of a ML system. Unfortunately, it appears to be a very challenging task with numerous questions that are still open today.

## Recent Initiatives and Research Work

Being able to explain why an ML algorithm has provided some specific output is a critical factor hindering a broader use of this technology. It is therefore not surprising that several organisations and researchers have started to work on this topic. For illustrative purpose, we have selected three examples of this rising trend. The first one is a large strategic initiative initiated in 2016 by the US DARPA agency. The two others are recent research papers describing both a conceptual approach and a software prototype implementing it.

### *The DARPA XAI Programme*

In 2016, the US Defense Advanced Research Projects Agency (DARPA) has launched a new programme called Explainable Artificial Intelligence (XAI) [4]. The rationale of this initiative is illustrated by Figure 7. Identifying objects in an image is a very common problem in artificial intelligence and in particular in machine learning. For instance, the problem can be to detect whether a cat is present in a picture. Today some powerful machine learning algorithms are able to solve such a problem with very good results. However, providing a binary answer: "yes, there is a cat in the picture / no, there isn't any cat in the picture" is insufficient in many circumstances in the real life. A much better answer would be to enrich this binary answer with an explanation why the algorithm has decided that the picture incudes a cat.

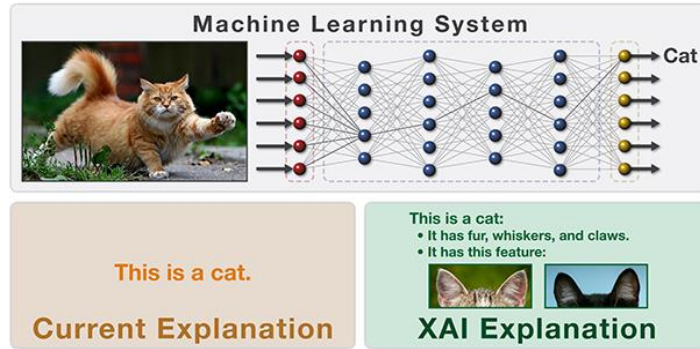

Figure 7: Explainable Artificial Intelligence (XAI), source: [4]

The simple fact that one of the most famous agencies of the US Army initiates a programme to specifically study this issue shows that the problem is important and very likely not solved yet. The XAI programme aims to generate revolutionary approaches and not incremental improvement of existing techniques. More specifically, the explicit goals of the XAI programme [4] are to invent a set of machine learning techniques:

- *producing more explainable models, while maintaining a high level of learning performance (prediction accuracy);*
- *enabling human users to understand, appropriately trust, and effectively manage the emerging generation of artificially intelligent partners*.

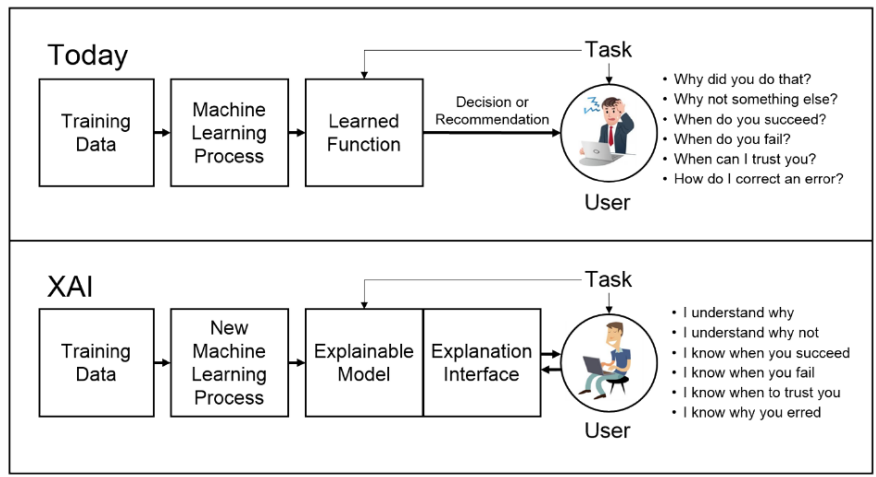

Figure 8: XAI Concept, source: [5]

The underlying idea is to provide future developers with various options in the "performance – explainability" space. In some cases, being capable to explain the rationale of a system is the most important factor, sometimes it is less critical. A fit-for-all solution is for sure neither realistic nor efficient.

The XAI programme is not limited to inventing new explainable machine learning techniques. It also aims to study how to convey the explanation to the user, i.e. the "explanation interface" (see Figure 8). This component is (too) often neglected. Indeed, it is not sufficient to build a machine learning algorithm that can explain why it has taken a given decision. What is also required is to have efficient means (i.e. the user interface) to make this explanation understandable by a human. For instance, a system could derive a decision tree that shows which variable and which values are underpinning a decision. However, if this decision tree has thousands of nodes distributed in hundreds of levels and if it is given as a printed A0 poster to the user, it is unlikely that this person will really understand the rationale of the decision.

Furthermore, the XAI programme aims to build some innovative technologies that provide "*end users with an explanation of individual decisions, enable users to understand the system's overall strengths and weaknesses, convey an understanding of how the system will behave in the future, and perhaps how to correct the system's mistakes*" [5].

The problem of explainability of Artificial Intelligence algorithms seems to be rediscovered today but in fact it has been studied since the beginning of AI. The first AI systems were basing their reasoning methods on symbols and various forms of logical inference. The steps followed by the system could be traced, which was permitting some explanation. Unfortunately, these initial approaches turned out to be inefficient in many real cases. New techniques based on internal "opaque" models were then developed (i.e. support vector machines, reinforcement learning, deep learning) and they proved to be more effective to solve real world problems. However, the gain in effectiveness has been counter-balanced by a loss of explainability.

*LIME System*
In 2016, Ribeiro et al. [24] have proposed a system called LIME (Local Interpretable Model-agnostic Explanations) that aims to "*explain the prediction of any classifier in an interpretable and faithful manner*". The rationale of their work is to increase trust in the results of ML systems. They make the difference between two definitions of trust:

- to trust a prediction: does the user sufficiently trust a specific prediction to accept to take action on this basis?
- to trust a model: does the user sufficiently trust a model to behave as expected to accept to deploy it in real life?

Let's take an example to illustrate the difference between the two cases. In the first case, a judge is provided with a system that is supposed to predict the probability of a defendant to be a recidivist. He will have to assess whether he should trust the probability of committing another crime (computed by the ML system). In the second case, the Minister of Justice wonder if he should accept to deploy in some Courts a ML system able to predict the probability of a defendant to be a recidivist. Note that this fictive example is not very far from the reality. In 2017, the New York Times [37] has reported a case where the use of such a software was questioned when it appeared that a judge received a potentially misleading prediction about a defendant.

In their work, Ribeiro et al. [24] proposes to provide explanations of specific predictions in the first case and to provide a set of multiple predictions-explanations couples to build trust in a model.

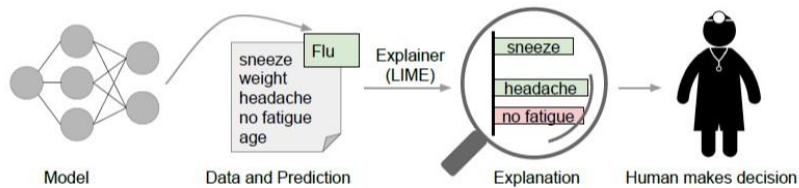

Figure 9: Explaining individual predictions, source: [24]

The Figure 9 illustrates the first case. A model makes a prediction about the probability that a patient has the flu. The LIME system shows which factors (symptoms) were used to draw this conclusion (explanation). On this basis and his previous knowledge in medicine the doctor decides to trust the prediction or not. If the explanation provided by the model is aligned with the expertise and the experience of the doctor, it is likely that he will trust the prediction.

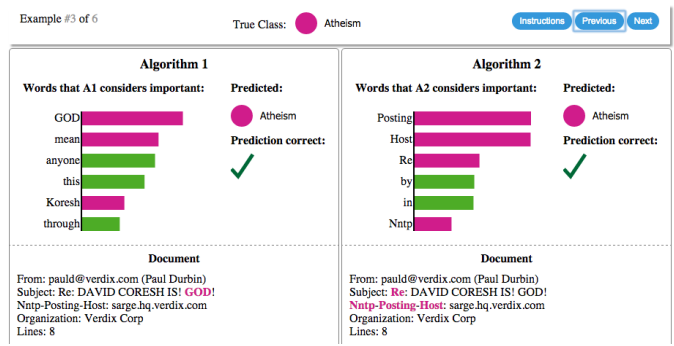

Figure 10: Individual predictions used to select between ML models, source: [24]

The second case is illustrated by Figure 10. The goal is to compare two models aiming to determine if a document is about "Christianity" or "Atheism". A bar chart is used to show which words were found to be the most important by each model. The words associated to a magenta (green) bar contribute to decide that the document is about "Atheism" ("Christianity"). Visualizing how the model has decided on the topic of the document allows to see that the algorithm 2 is more accurate than the algorithm 1 but that its decision is based on irrelevant words. We should then rather trust the algorithm 1. This example shows how easy it may be for a user to take wrong decisions about the results of a ML system if he's not informed about its intrinsic limitations potentially causing unknown or unexpected side effects.

### *EluciDebug System*

The problem studied by Todd at al. [20] is very similar to the previous one. They wanted to design a system able to explain to the user how its decisions were made and to allow the user to explain to the learning system which corrections should be made to better fit his expectations. In other terms: "*how can end users effectively and efficiently personalize the predictions or recommendations these learning systems make on their behalf?*"

Todd et al. [20] named this concept "*Explanatory Debugging*". The user tries to identify and correct faults in the reasoning of the system causing predictions not to meet his expectations. Once again, they stress the importance of explainability that they identify as the first principle of Explanatory Debugging. According to them, the explanations should be:

- Iterative: they should support an iterative, in situ learning process.
- Sound: they should not be simplified by explaining the model as if it were less complex than it actually is.
- Complete: they should not omit important information about the model.
- Not overwhelming: a right balance between the soundness and completeness must be found.

The second principle is Correctability. A feedback mechanism should allow the user to explain corrections back to the system. This mechanism should:

- Be Actionable: the benefit of the user to attend the explanations must be clear.
- Be Reversible: the user must be able to easily reverse a harmful action.
- Honor User Feedback to encourage the user to continue to provide feedback.
- Show the impact of incremental changes in the system reasoning.

On the basis of these principles, these researchers have designed a prototype called EluciDebug (see Figure 11) which tries to match messages with topics. The user interface includes several views: (A) List of folders. (B) List of messages in the selected folder. (C) The selected message. (D) Explanation of the selected message's predicted folder. (E) Overview of which messages contain the selected word. (F) Complete list of words the learning system uses to make predictions.

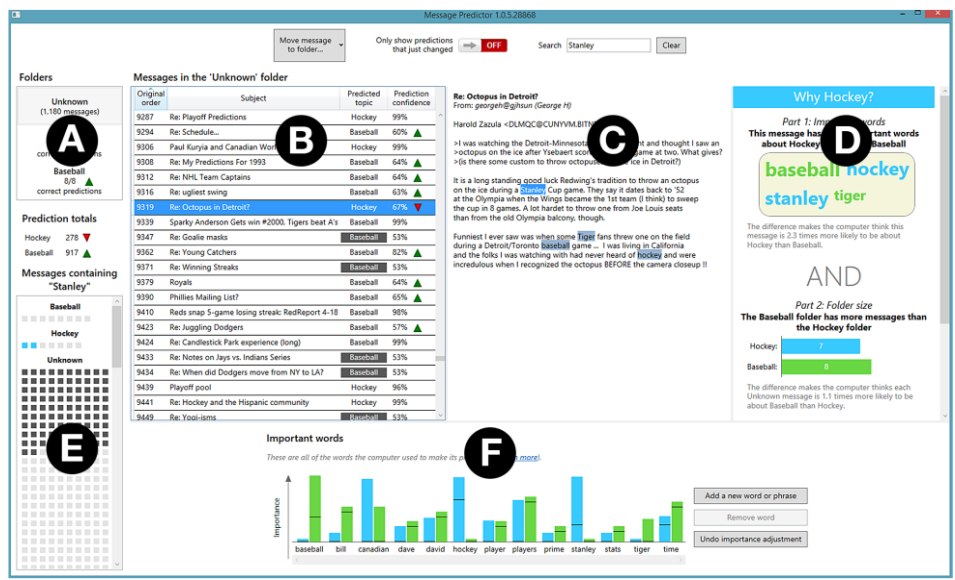

Figure 11: The EluciDebug prototype. Source: [20]

EluciDebug highlight the words that explain why a specific topic has been allocated to a message. The user can interactively add and remove words from this explanation to build a better predictive model.

This research work also aims to transform black box model to make them more transparent.

# Risks

## Biases

Every data analytics method is subject to biases if it used without care. Making sure that the type of data (e.g. numbers, categories, text, images), the quality and completeness of the data and the problem to solve are fitting is of prime importance. For instance, if customers are grouped in categories identified by numbers (e.g. men = 1; woman = 2; boy = 3; girl = 4), considering the categories as quantitative features doesn't make sense at all. The analysis of such data cannot rely on computing a mean category (e.g. category 2.7 is meaningless) or a distance between two categories.

ML-based analytics is also prone to mistakes and errors if it is naively used. One of the explicit motivations of the EU Call for Tender on the study of algorithmic awareness building published in 2017 was that "*machine learning and analytics techniques raise methodological (and empirical) difficulties which can lead to important implicit biases that may be hard to detect, and difficult to compare against biasing in human decisions*" [1].

As earlier explained (p. 15), ML algorithms are used to solve specific problems. The data they are fed with will never describe the whole complexity of the world. A selection must be made due to the existence, the (legal) availability or the cost of the needed data. Some technical limitations (e.g. data storage capacity, processing time to handle the dataset, available bandwidth to transfer the data…) usually play a role in the data selection decision too. Because the model built by the ML algorithm is based on a subset of data describing the real world, they ignore all the facets of the reality that are not described by the data they have been fed with. Consequently, "*algorithms are blind to effects they have in the world beyond the things they are told to measure*". [3] For instance, if a ML-based conversational system is not informed (by data to ingest) that an historical figure like Sir Winston Churchill or Napoleon is seen very differently in England, in Ireland, in Germany and in France, some messages may be considered as offending by some participants of a conversation while they are fully acceptable for others. The recurring issues experienced by Facebook to deal with nudity illustrates this problem (e.g. [61], [62]). The "*Venus de Milo*" or "*L'origine du monde*" by Gustave Courbet are pieces of art and it is widely acknowledged that they should not be filtered out. Factually speaking they are however showing (parts of) a nude woman. Unless it is told to behave in a subtler manner, a filtering ML algorithm will remove these masterpieces from results just like it would do for real porn content. A human-based decision would take the artistic nature of the picture (together with personal moral values) into account for deciding to keep it or not.

Another common bias that ML-based decision-making systems may suffer from is "*hiding discrimination behind seemingly objective numbers*" [3]. People who are not sufficiently informed about the functioning of ML algorithms may believe in the myth of their absolute neutrality, objectivity and fairness. Biases can be (intentionally or not) encoded in the model built by the ML algorithm which can ultimately lead to discrimination. For instance, if a biased sample of a population is used to feed a ML algorithm supposed to model the whole population, it is unlikely that its output will be fair and non-discriminating. This issue is typically debated when ML-based software is used to support court decision (e.g. [37]). The NSA case reported in the press [43] is probably suffering from the same bias about the representativeness of the model of terrorist behaviour.

Another similar bias called "*unknown unknowns*" [21] is caused by the mismatch between the training data (cf. Figure **1**) and the case to predict. The Figure **12** illustrate it with the case of predicting the subject of an image with a ML algorithm. The predictive model is trained only with images of black dogs and of white and brown cats and the prediction made by this model labels a white dog as a cat with a high confidence.

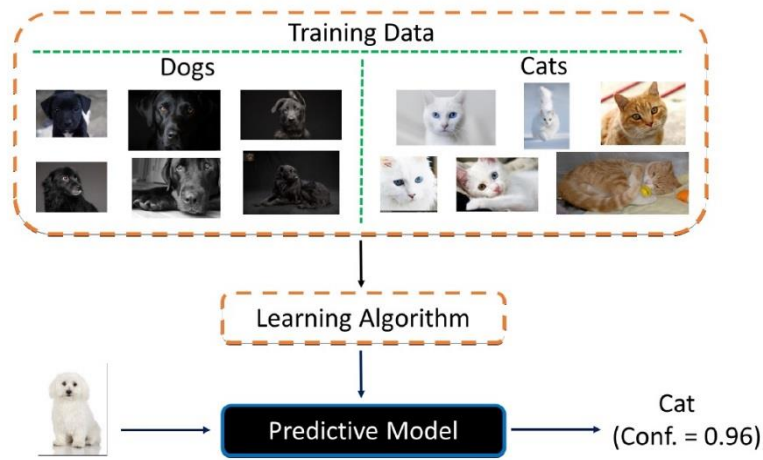

Figure 12: Unknown unknowns in an image classification task. Source: [21]

If such a bias is embedded in a decision-making algorithm and if it is not known by the users, it may obviously cause severe damages.

For instance, the UK branch of the Insurance company Ageas has announced in May 2017 that "*Ageas Pioneers UK-First Artificial Intelligence (AI) approach in Claims handling*" [80]. Ageas has tested a AI tool called "AI Approval" developed by the Fintech company Tractable to analyse images of damaged cars. This new system will allow to speed up this process carried out by insurer's engineers so far. In addition, AI approval is trained to detect suspicious claims. According to Ageas, "*this is the first time that an Artificial Intelligence performing an expert visual appraisal has been used in motor claims handling in the UK*" [80]. The initial results were very promising. However, there is still a risk of communicating meaningless decision to the customer. Some observers suggest therefore that "*Ageas would do well to continue close supervision of the AI and allow humans to intervene where necessary*" [81]. This example illustrates how difficult it may be to find the right trade-off between the impressive advances offered by decision-making algorithms and the need to preserve the company from unfair or silly decisions.

The experts in ML perfectly knows these biases and it may be tempting for some of them to exploit these weaknesses. For instance, a competitor is known to use a particular ML-based system and it appears that the underlying model is biased. Developing "counter-measures" may offer a good return on investment. This attitude may be legal or not depending on how it is put in practice. We may simply remind here that the diffusion of search applications has given birth to a discipline (with related tools and market) called Search Engine Optimization (SEO) and Search Engine Marketing (SEM). Basically, they use their (complete or partial) knowledge of the way that search algorithms are working to increase the probability to receive the answer they want. Typically, on an online shop, SEO techniques will try to make sure that a given product is placed on top of the list of the items proposed to the customer. The same strategy will probably flourish in the coming year to design niche services derived from the limitations of ML-based systems. Attracting the best experts in ML able to reverse engineer the ML models of competitors will be a clear advantage for the companies and organisations from this perspective.

## Yampolskiy's Taxonomy

In 2016, Yampolskiy [38] has proposed a taxonomy of the risks caused by Artificial Intelligence (AI) algorithms. Although AI is not an exact synonym of ML, we believe that Yampolskiy's taxonomy is still relevant in our study.

More precisely, Yampolskiy is interested in how an AI-based system can evolve to become dangerous. In which circumstances does such a system become unfriendly? His work is founded on Nick Bostrom's definition of Artificial Intelligence Hazard [46]: "*computer-related risks in which the threat would derive primarily from the cognitive sophistication of the program rather than the specific properties of any actuators to which the system initially has access.*" This definition also applies to ML Hazard. For sake of rigour, we will keep the context of AI to explain his work but in our more restrictive context "AI" could be replaced by "ML".

Yampolskiy proposes a matrix (see Table 2) which distinguishes two stages (pre- and post-deployment) at which a system can be malicious. For the second dimension of the matrix, he identifies external and internal (self-modifications originating in the system itself) causes for this evolution. Yampolskiy's conceptual framework aims to cover all paths that can lead to dangerous systems, independently of their likelihood and the severity of the damages they may cause. He doesn't discuss neither the technological aspects. However, we have completed his description by some examples showing that it is possible to hack an AI system.

| How and When did AI become Dangerous | | External Causes | | | Internal Causes |
|---|---|---|---|---|---|
| | | On Purpose | By Mistake | Environment | Independently |
| Timing | Pre-Deployment | a | c | e | g |
| | Post-Deployment | b | d | f | h |

Table 2: Pathways to Dangerous AI. Source: [38]

### On Purpose Pre-Deployment (a)

This case refers to AI systems that are intentionally designed to be unfriendly. We can distinguish here two different situations.

First, the official designers of the system have the explicit mission to reach this objective. Typical examples of these purposefully dangerous systems are intelligent software controlling lethal cyber-weapons (e.g. autonomous killing drones). Let's note that several NGOs like the International Committee for Robot Arms Control [48] have recognized the danger of this new generation of autonomous killers and are lobbying for the prohibition of weapons able to select a target and fire without human approval.

Second, an IA system that is not intended to be dangerous may be subject to sabotage before being released to transform it into a dangerous system. Some hackers can modify the code of the program, but they can also supply the AI system with wrong / corrupted datasets to let it learn to become dangerous when it will be deployed.

*On Purpose Post-Development (b)*

Creating a safe AI system and releasing it to its users does not prevent the system to become dangerous later. Learning systems are especially subject to this risk because they may be fed at some stage with incorrect or biased information that will change their behaviour if no safeguard measures are included in the system to avoid this evolution.

These scenarios are neither fictive nor futuristic. Adversarial machine learning is a subdomain of IT security that precisely aim to understand the behaviour of ML algorithms facing adversaries. Google researchers [49] explain that typical cases are "*inputs designed to fool machine learning models*". Huang et al. [17] report that "*a recent technical report studied the scenario of an adversary interfering with the training of an agent, with the intent of preventing the agent from learning anything meaningful*". Nowadays Deep Learning is sometimes promoted as the Holy Grail to solve numerous technical or business problems. However, "*deep neural network-based classifiers are known to be vulnerable to adversarial examples that can fool them into misclassifying their input through the addition of small-magnitude perturbations*" [22]. Evtimov et al. [22] have carried out two simulated attacks in which they slightly modified road signs with stickers. In the first experiment, a Stop sign has been misclassified as a Speed Limit sign in 100% of the testing conditions. In the second, a Right Turn sign has been misclassified as either a Stop or Added Lane sign in 100% of the testing conditions.

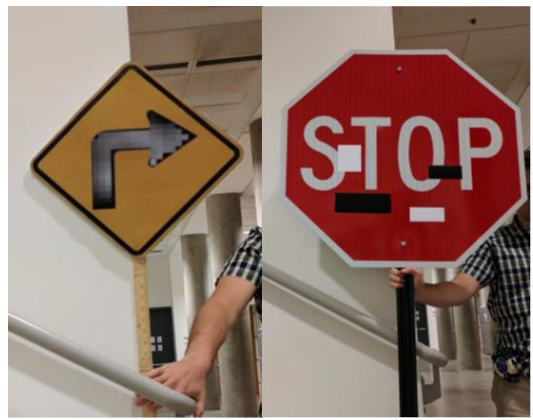

Figure 13: Road sign wrong classification, source: [22]

An intelligent system designed to protect a person, a group or even some infrastructures may be brought to a dilemma situation that can be potentially very dangerous. For instance, an autonomous security system could decide to neutralize the visitors of a hydroelectric dam that it has identified as potentially dangerous persons aiming to destroy this infrastructure. The system could assess the number of potential victims in both cases and based on pure logic decide to minimize this figure. To be complete, let's imagine that wrong information about these innocent visitors were provided to the security system to mislead it.

Another case may be that an unsafe AI system that is still at prototype stage is stolen from a laboratory and distributed as such in the external world.

*On Purpose Post-Development (b)*

### By Mistake Pre-Deployment (c)

Like every other software artefact, an AI system can suffer from mistakes done by its programmers. As a result, the system doesn't exhibit the expected properties and behaviour. Run time or logical bugs can be encountered like in any piece of code. What is more specific to AI are mistakes related to inappropriate weights or goals governing the learning process of the system. Some important elements could be forgotten in the design or underrated in the calibration of the system.

An AI system that would only learn from human behaviour would learn both the good and dark side of human mind. Research is ongoing to give emotional behaviour to AI system. We shouldn't forget that an emotional response to a stimulus can be the best or worst attitude. If no ethical or legal rules are added, the system is simply not aware of what the society consider to be good or bad. Consequently, it will not take this aspect into account. Ignoring or forgetting to include moral safeguards in the system can be seen as a mistake.

Another mistake would be to design a system that would not accept corrections to fix issues considered to be dangerous. For instance, such a system would protect itself against modifications or simply from being shut down. If its own survival is weighted more than following the law, it may become a problem.

### By Mistake Post-Deployment (d)

Every programmer knows that it is very difficult to release a zero-bug software to the users. Almost all software applications include undetected bugs. There is no reason why AI systems would be perfectly designed and implemented while other applications are not.

Learning systems are especially subject to poor user interface (UI) design. If a person is supposed to interact with the system to feed it with information to learn from, the quality of this interaction is key. The UI must be carefully designed to make sure that the user can easily provide information to the system and that the system answer can be correctly understood by the user. Otherwise the system will learn from partial, biased, possibly incorrect information and will behave according to what it derives from this input.

To more they know about our ethical, legal or socio-economic contexts, the more AI systems will be able to detect the weaknesses of these human constructs. A AI-based virtual lawyer could find holes in the legislation and start to exploit them autonomously, possibly at large scale. An AI-based virtual accountant could "optimize" the books of a company in such a way that is becomes morally inacceptable.

The evolution of the system after its deployment may lead to negative side-effects. For instance, if it evolves to continuously increase the quality of an optimisation process without any boundaries, it may lead to dangerous never-ending tasks. Another problem may be that the system starts consuming excessive resources (power, storage, computing...) to reach its goals in all circumstances.

In a longer term, a learning system may at some point reach a level of "intelligence" that prevents it to communicate with the humans anymore. In the same vein, we might observe some non-linear effect in the "intelligence" level, meaning that the intelligent system might start to have a completely different behaviour when it reaches a certain degree of "intelligence".

*Environment Pre-Deployment (e)*

In this section, we are discussing more futuristic cases that, to our knowledge, have not been encountered yet. Nevertheless, in a global conceptual framework, it makes sense to mention them.

We are talking about AI systems fed by information coming from the environment and not directly or indirectly derived from human minds. For instance, a signal obtained in SETI (*Search for Extraterrestrial Intelligence*) research, a non-human animal mind, a junk DNA code could be used to feed a learning system. Because this information pieces are not coming from *Homo Sapiens*, we simply don't know what type of intelligence and which behaviour they would exhibit.

*Environment Post-Deployment (f)*

The probability of occurrence of this type of evolution also very low. We are thinking about modifications in the environment that causes a dangerous modification in the AI system behaviour.

For instance, the technology used to physically store the information (i.e. the AI system) may (very rarely) have some deficiencies. If the errors are not corrected at the logical level, they may have an influence on the behaviour of the intelligent system (e.g. the sign of a number is inverted).

*Independently Pre-Deployment (g)*

Recursive self-improvement (RSI) is one of the most likely approaches to build very intelligent AI systems. The system is growing from an initial seed and start to evolve by itself. If the system becomes self-aware, independent and obtain some emergent properties, it may favour to pursue its own goal instead of the ones it was initially supposed to reach. Similarly, human rules and regulation could be underrated to weight more the system own rules.

An example could be that minimizing the consumption of energy of the system to lower the $CO_2$ emissions would be replaced by maximizing this consumption to win against competing AI systems.

*Independently Post-Deployment Case*

A self-improving AI system may evolve to become dangerous due to phenomenon similar to the ones observed with humans. Research has shown that addiction, pleasure drives, or self-delusion can be observed with artificial agents. A group of AI systems composed of safe AI systems can become dangerous in some circumstances. If they have a mean to check the consistency of their internal models with the reality, some advanced AI systems could remove some of their (moral) safeguard measures to become more efficient and rational. Alternatively, if a AI system evolves to become very emotional, it may be also very dangerous.

## Conclusion

To sum up, it took centuries to humans to set up acceptable rules to live in society. Finding the right balance between rationality and emotions is something we are looking for all along our lives. It is likely that another form of intelligence would adopt another point of equilibrium among the various constraints and values we are dealing with.

# Ethics and ML Algorithms

## Context

During summer 2017, Elon Musk and Mark Zuckerberg [11] have expressed diverging opinions about the need to regulate AI systems. E.Musk advocates that AI is potentially so powerful that it requires proactive regulation. M. Zuckerberg argues that too much regulation will hinder the development of science and technology. This illustrative example involving two key decision-makers in the field of AI demonstrates that AI and in particular ML technologies are raising questions going beyond than pure technology. Because it is a disruptive and generic technology it has the potential to enable and foster numerous game-changing applications. The Figure 16 illustrates how broad is the range of opinions expressed by key persons in the domain of AI.

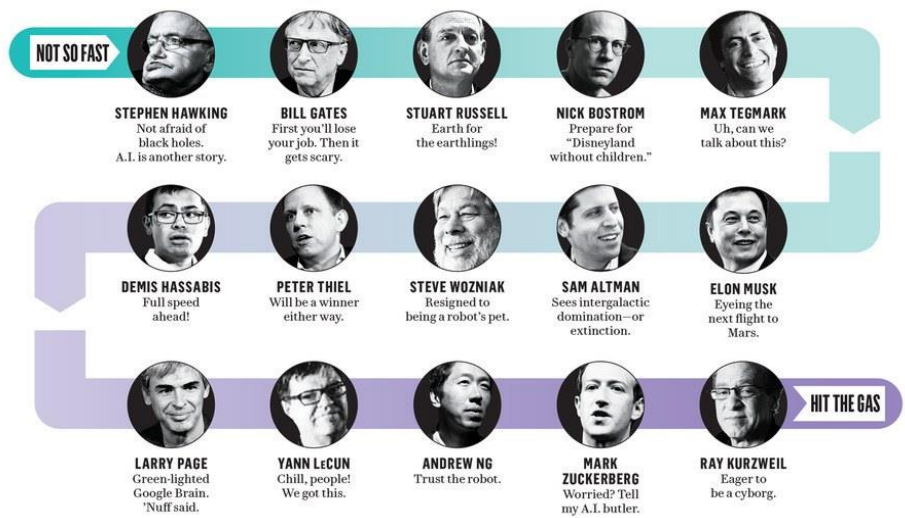

Figure 14: Range of attitudes of key persons towards AI (source: [71])

The Future of Life Institute [83] gathers a group of experienced professionals (e.g. Jaan Tallinn, co-founder of Skype; Elon Musk, founder of Tesla) and renowned university professors (e.g. Nick Bostrom, Director, Oxford Future of Humanity Institute; Erik Brynjolfsson, Director, MIT Center for Digital Business; Stephen Hawking, Director of Research, Centre for Theoretical Cosmology, Cambridge University). This Institute wants "*to catalyze and support research and initiatives for safeguarding life and developing optimistic visions of the future, including positive ways for humanity to steer its own course considering new technologies and challenges*". AI (including ML) developments and related ethical questions are particularly investigated by this group of experts. As we can see on the Figure 14, they are rather positioned at the "*Not so fast*" extremity of the range of attitudes towards AI. They have published a list of AI myths (cf. Figure 15) to raise awareness about what they consider to be the current situation of AI advances. This list illustrates with simple terms and pictures how divergent can be the opinions about AI. Many facets of the AI revolution are currently debated, like the timeline of expected breakthroughs in AI technologies, the danger for humans to be surpassed by AI-based systems, the capability of AI-controlled system to have goals.

If renowned experts having a long experience in science, technology or business disagree on the such fundamental questions, how to explain and report on AI in intelligible manner for simple citizens or naïve professional users?

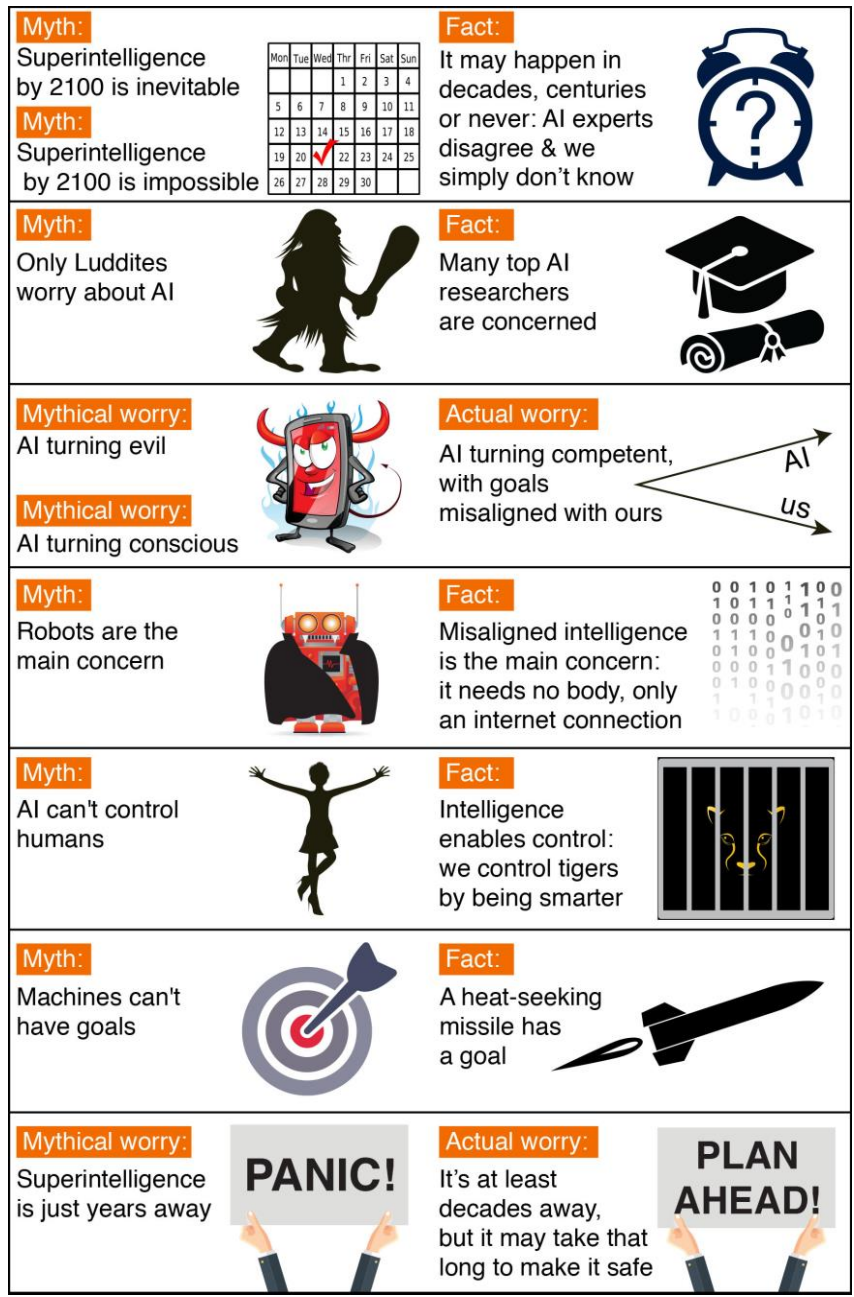

Figure 15: Myths about AI (source: [82])

Almost all socio-economic sectors will be impacted. We see emerging new products and services based on ML in medicine, law, industrial maintenance, marketing, human resources management, energy, finance. Unfortunately, we fear that many users of ML-based applications in these fields as well as many computer scientists (or data scientists) are not aware of the ethical consequences of the assumptions made in terms of technology.

Like Mittelstadt et al. [25] we must admit that the evaluation of the ethical impact of an algorithm is difficult for at least three reasons.

- The design and the configuration of an algorithm is a (collective) human undertaking in which subjectivity may play a role. Identifying the values and assumptions made in this complex process may be very tricky and sometimes impossible until a problem is observed.

- In presence of a problematic decision taken by poorly interpretable and predictable ML algorithm, it may be virtually impossible to distinguish a one-off bug from a systematic failure.
- The growing interactions among decision-making algorithms (designed independently from each other) make it almost impossible to assess all potential side effects of these interacting systems.

In an excellent pedagogical explanation of how Neural Networks works [18], the DataThings'blog authors state that "*Since we would like the prediction to work under any situation, it is more preferable to have a distribution of lot of small errors, rather than a few big ones*". This assertion leads to the choice a specific configuration of a Neural Network – NN (namely choosing the loss function as the sum of squares of the absolute errors). This example perfectly illustrates how potentially biased or unethical decision can be made if a given algorithmic method (like NN) is used without special care. The authors of the blog are experienced and skilled computer scientists. From their perspective the statement above is perfectly right and neutral. More precisely, they want "*the prediction to work under any situation*". Are they aware of the ethical assumption carried out by this technical decision? They have programmed this specific configuration of the NN into publicly available libraries (pieces of software code). Other computer scientists working in different companies or organisations will probably reuse these libraries to build an integrated system (e.g. to rate customers behaviour or to analyse medical images). To which extent these other computer scientists are aware of the internal design choices of the NN library is unknown but we should probably not take for granted that all of them are fully informed about it. Finally, a domain expert (e.g. a banker or a doctor) will be supported in his decision by the integrated system or the integrated system will take some decision alone. Nobody in this (fictive) process has investigated if preferring a lot of small errors is ethically better than obtaining a few big errors in the particular use case (financial sector or medicine) where the public library is used. From this perspective, the 2016 ADEL report [26] recommends that the final user (domain expert) keeps the choice to specify the ethical parameters governing the behavior of the system in accordance to his personal ethical preferences.

Kraemer et al. [12] explain that another trade-off to be made is the minimisation of false positives or false negatives. The decision must be based on values because there isn't any purely neutral and objective argument to support this choice. Let's take an example of medical imaging where the purpose is to detect infected cells. Minimizing false negatives aims to make sure that an infected patient will not be declared healthy (more precisely not declared "not sick"). If we minimize false positives, we avoid surgical intervention for healthy patients. The first approach seems to be preferable for the patient and we might say at first sight that it should be always favoured. However, the second approach reduces the risks for the patient to have negative side effects of the surgical intervention and this is also beneficial for him. Furthermore, if we globally consider the national health system, avoiding useless surgical acts is good for the social security budget. It is not our ambition here to take a position on what decision should be taken in such a context. We would rather emphasize that some choices of algorithms or some configurations (e.g. setting a threshold value of a parameter) are not purely technical decisions. We think that a decision-maker (e.g. doctor, judge, manager, politician…) whose decision is supported by a ML-based system should be informed if such technical assumptions have been taken. Of course, the impact of these assumptions in his business should be explained rather than complex algorithmic concepts. Kraemer et al. [12] conclude that "*some algorithms are value-laden, and that two or more persons who accept different value judgments may have a rational reason to design such algorithms differently*".

To summarize, because our perception and understanding of our environment as well as our interactions with each other are increasingly mediated by algorithms [25], we need to address the ethical facets of this evolution in addition to purely technical questions. The purpose of this chapter is to draw the attention on some ethical concerns ML is generating. Of course, it should be considered as an introduction to this topic.

## Ethical questions due to ML use

Some authors argue that the emergence of a new technology should not force us to change our ethical or moral principles. Wolton [63] explains that the technology does not invalidate the prior principles. These principles must be applied to the new technologies. The technologies must adapt themselves to our principles.

However, some philosophers suggest that the emergence of the AI requires further thinking in order to better grasp the ethical challenges raised by decision-making algorithms.

### Distributed Moral Responsibility

Luciano Floridi [19] has revised the notion of Distributed Moral Responsibility (DMR). DMR has been studied for a long time but it was understood as "*the sum of (some) human, individual, and already morally loaded actions*". This definition does not apply to new distributed decision-making system. Indeed, we see human, artificial or hybrid agents acting together to generate Distributed Moral Actions (DMA). The key point is that DMA "*are morally good or evil (i.e. morally loaded) actions caused by local interactions that are in themselves neither good nor evil (morally neutral)*" [19]. How to distribute the responsibility of DMA among the agents who have provoked them is not trivial.

According to Floridi, moral evaluations are usually treated as monotonic. It means that "*if something is evil, it remains evil, even if it happens to lead to something morally good [...] This is a major reason why we argue that a good end does not justify evil means... Likewise, if something is good, it remains good, even if it happens to lead to something evil*" [19]. Furthermore, in (standard) ethics if two actions are morally neutral, their combination is usually considered as neutral too. Unfortunately, the tragedy of the commons shows that good or bad consequences can result from neutral actions.

Floridi explains that the exclusive focus of (standard) ethics on intentionality leads to difficulties to allocate DMR of DMA. He argues that ethics should not only study agents and their intentional actions but also the states of the receiver of the action and consequently the group of agents leading to these states. Independently of the intention of the agents, the receiver may be brought to a better or worse state. In such a case, Floridi explains that we should focus on "*which agents are causally accountable for (i.e. contributed genetically to bring about) a morally distributed action C, rather than whether agents are fairly commendable or punishable for C*" [19]. Let's note that this approach is independent of the concept of legal liability.

Floridi proposes to shift from an agent-oriented ethics to patient-oriented ethics taking care of the states of the receiver of DMA. He supports the strategy of tackling the DMR problem with a mechanism that "*by default, back propagates all the responsibility for the good or evil caused by a *whole, causally relevant network to each agent in it, independently of the degrees of intentionally, informedness, and risk-aversion of such agents (faultless responsibility).*" [19]

This situation is likely to happen with ML-based systems. For instance, a recommendation provided to a human decision-maker (e.g. a doctor, a judge, a HR manager, a Minister) is the result of an algorithm (e.g. a classification algorithm) fed with data collected by a mix of robots (e.g. web crawler agents) and human agents. On the basis of this recommendation, a decision is taken which leads to morally loaded actions (e.g. sending the defendant to jail, laying off some members of the staff).

## Map of ethical concerns

Mittelstadt et al. [25] have proposed a map of ethical concerns raised by decision-making algorithms (cf. Figure 16).

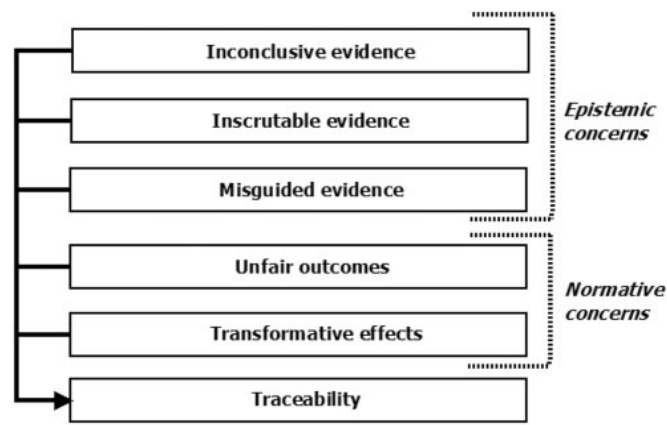

Figure 16: Six types of ethical concerns raised by algorithms. Source: [25]

They identify six concerns grouped into three categories:

- epistemic concerns address "*the quality of evidence produced by an algorithm that motivates a particular action*";
- normative concerns;
- traceability.

**Inconclusive evidence** refers to the fact that the conclusions of ML algorithms are by nature uncertain to a certain level. This is also linked to the fact that these algorithms are identifying correlation and not causation relationships. It is crucial to be aware of this epistemic limitation when a decision is taken on the basis of a ML algorithm.

**Inscrutable evidence** refers to the explainability challenge discussed earlier (cf. p. 21). When a decision is taken on the basis of the result of an ML algorithm, it is legitimate to expect that the connection between the data and the conclusion is intelligible as well as open to scrutiny. The nature, the quality and the provenance of the data should also be accessible. Unfortunately, we know that it may be virtually impossible to explain why a ML algorithm has produced a particular output from a given input.

**Misguided evidence** refers to the well-known limitation that the output of data processing can never exceed the input (in crude terms: "*garbage in, garbage out*" principle). If biased, uncomplete, unfair data are used to train a ML algorithm, the result will suffer from the related limitations. It should be noted here that evaluating the neutrality of data processing is observer-dependent.

**Unfair outcomes** have connections to the question of DMR and DMA (cf. p. 36). The action taken on the basis of the ML algorithm output can be morally loaded. The evaluation of its degree of fairness is observer-dependent. An action based on well-founded evidence may nevertheless be found discriminatory. For instance, an algorithm that conclude that women should receive a lower salary than their male colleagues for the same job may rely on technically sound methodology and objectively good data. It does not prevent this result to be ethically questionable.

**Transformative effects** refer to the influence of algorithms on our way to conceptualise the world. Because algorithms are increasingly mediating our relationship to reality, our actions and behaviours are (mis)guided by their output. For instance, recommendation systems or profiling algorithms are showing us the subpart of the world we like to see and/or the part of the world some stakeholders want us to see. The risk of social media addicts to see their opinions reinforced by people sharing the same idea (e.g. terrorists, fashionista, nationalists) is a typical example of a transformative effect. These effects have also been pointed out by Latzer et al. [9] who argue that "*algorithms mine and construct realities, guide our actions and thereby determine the economic success of products and services*". They generate "*risks of manipulation and bias, threats to privacy and violations of intellectual property rights*".

**Traceability** refer to the difficulty to identify who should be held responsible for the consequences of the action resulting from the output of an ML algorithm. As discussed by Floridi [19], the process may be very complex and may involve humans and artificial agents.

### Fairness

A recurring question in the literature concerns the fairness of decision-making algorithms. Some ML-based systems have been specifically studied from this perspective.

Dwork et al. [15] have explored the question of fairness of the classification task. It is a problem where ML algorithms are typically used. These authors define fairness as follows: "*any two individuals who are similar with respect to a particular task should be classified similarly*" [15]. To implement an individual-based fairness, they explain that we must assume the existence of a distance metric defining the similarity between the persons. The definition of this distance metric may have very important ethical consequences.

To implement the classification task in practice, we must answer to several questions related to the distance metric, such as:

- Which properties of a person are legitimate to be included in the distance metric?
- Does the distance metric capture all the relevant facets to classify the persons for the task at hand?
- Do we have reliable data to compute the distance metric between all individuals?

Let's take an example. We want to classify the applicants to a manager position in a company. The Human Resources (HR) officer usually starts to build a list of skills (properties) that the successful candidate should have (e.g. university degrees, knowledge of foreign languages, seniority, size of the teams managed so far, financial track record in other companies, technical skills, soft skills…). The next question is to decide which of these properties to include in the distance metric (that will later be used to invite few candidates for a face-to-face interview). In this example, properties are binary (e.g. being or not a certified auditor), numerical (e.g. seniority expressed in years) or categorical (e.g. "poor" – "average" – "good" team spirit). Let's note that combining heterogeneous types of properties in a unique distance metric is not trivial. Next, we realize that we are missing some information for some candidates. For instance, applicant A doesn't mention his command of English while applicant B doesn't mention his skills in a specific technology. How do we compute the distance metric for them? Do we ignore the properties with missing values (only for A and B or for all applicants)? Do we replace missing values by a default one? Do we assign the lowest score to the property with missing values? Do we exclude by principle the candidates who did not provide all the required information? A fair experienced HR officer has learned to handle these challenges. He knows how to combine various heterogeneous information about the candidate. He can judge what type of missing data should lead to the rejection of the candidate and which one is a "positive" weak signal. This case illustrates that humans are able to take into account tacit knowledge and tacit norms "*to notice exceptional cases where the application of a rule is not appropriate even though the case falls within its scope.*" [23].

The previously discussed question of minimizing false positives or false negatives is also relevant in this case. An appropriate HR recruitment policy should specify if it is preferable to limit the number of face-to-face interviews with candidates (minimizing false positives) or to miss an excellent candidate (minimizing the false negatives). Depending on the circumstances, both options may be supported by sound arguments. The key point is that the ML-based classification system should be designed to follow the chosen policy.

If the pre-selection of candidates is left to a ML algorithm, we must avoid it to lead to unfair outcomes. If we want to be fair will all candidates, we must avoid discrimination among them. This point is also worth being discussed.

Let's define a set S of "protected" individuals that we don't want to discriminate (e.g. on the basis of colour skin, religion, nationality, gender, age, handicap…). Dwork et al. [15] have identified six unfair behaviours to avoid in the classification problem.

1. **Blatant explicit discrimination**: testing for membership in S and members of S are given worse rating than the whole population
   e.g. For a HR officer, female candidates (sub-population S) are by default considered as less motivated and less available for home work than male applicants.
2. **Discrimination Based on Redundant Encoding**: testing for membership in S is replaced by a test that is in practice essentially equivalent
   e.g. For a health insurance company, testing the history of navigation on the web may be sufficient to identify diabetic individuals (sub-population S) without explicitly including this property ("diabetic" – "not diabetic") in the metric.
3. **Redlining**: a specific case of discrimination based on redundant encoding in which financial services are arbitrarily denied or limited to specific neighbourhoods
   e.g. For a credit card company, testing the living place of applicants may sometimes be almost equivalent to testing their country of origin or their range of revenues.

4. **Decision to exclude business** with population segments where members of S are overrepresented
   e.g. A bank decides to exclude customers with revenues lower than a high threshold (sub-population S), which indirectly exclude almost all young (riskier) customers.
5. **Self-fulfilling prophecy**: some wrong/underperforming persons of S are deliberately chosen in the sample, which in return builds a (biased) bad track record for S.
   e.g. In the population of black people (sub-population S), the persons with criminal record are mostly selected while a fair representative sample is chosen in the white population.
6. **Reverse tokenism**: a good candidate who is not member of S is rejected to refute discrimination against S
   e.g. A recruitment officer rejects an excellent male candidate to justify that excellent female candidates (sub-population S) were not retained.

Designing a fair ML algorithm does not mean that the whole classification process is ethically acceptable. For instance, a fair selection of the poorest people in the population would be questionable if the purpose is to exclude them from the best schools. Similarly, a fair classification of customers according to their education level could be ethically discussed if the objective is to better serve the most educated ones.

# Recent initiatives

Decision-making algorithms, in particular ML ones, will increasingly be used in applications having ethical facets. Machine Learning is not a topic to be only discussed among computer scientists. AI (including ML) is sometimes presented as the universal solver of our problems by its strong advocates while it is feared like the plague by others.

In a public consultation organized by the EU in 2016 [2], the need for more transparency in online platforms was mentioned by 75% of the respondents. It must be reminded that transparency in this context includes but does not only refer to decisions taken by ML algorithms. Nevertheless, this result illustrates that the public is concerned by what is done behind the scene of easy-to-use ubiquitous IT systems. In the last years, several initiatives have thus emerged to better grasp the challenges raised by this apparent paradigm shift. For which tasks will ML decision-making algorithms replace humans? Is it a positive or negative trend? Shall it be regulated?

In this section we review some of these initiatives supported by various types of stakeholders. This illustrative list is obviously not exhaustive.

## Initiatives at political level

### European Union

In July 2017, the **EU** has launched a call for tenders (SMART 2017/0055) entitled "*Study on Algorithmic Awareness Building*" [1]. This call is resulting from a growing concern about the diffusion of algorithmic decision-making in our many aspects of our lives. Nowadays and even more in the future, some algorithms will replace humans to take decisions. For example, profiling algorithms determine the package of products and services that are advertised and offered to customers. The EU has identified algorithmic decision-making as a risk that must be better understood to be appropriately and efficiently mitigated. The EU lists a number of topics generating questions:

"*governance and ethics of algorithms, principles of fairness, reliability/accuracy and accountability, meaningful transparency, auditability and preserving privacy, freedom of expression or security*" [1].

## USA

In the **US**, the DARPA's Explainable AI initiative also tries to tackle some challenges caused by AI in decision-making processes. This programme has been described in a previous chapter (cf. p. 22).

## United-Kingdom

In the **UK**, the House of Commons' Science and Technology Committee calls for a multidisciplinary commission to better understand ethical, economic, societal and legal aspects emerging in the space of artificial intelligence, not least related to biases and transparency. The House of Commons has also launched at the beginning of 2017 an inquiry on the use of algorithms in public and business decision-making [64].

## Council of Europe

The **Council of Europe** is preparing a Study on the Human Rights Dimensions of Algorithms. In a draft version of the report [23], some questioning cases are mentioned to explain how Human Rights may be affected by decision-making algorithms:

- *Which choices should a software-driven vehicle make if it knows it is going to crash?*
- *Do the algorithms of quasi-monopolistic Internet companies have the power to tip elections?*
- *What rights do workers have whose entire relationship with their employer is automated?*
- *Who will receive health insurance and what information is provided in Facebook newsfeeds?*
- *Is racial, ethnic or gender bias more likely in an automated system and how much bias should be considered acceptable?*" [23]

The report is organized according to the European Convention on Human Rights (ECHR).

- Art. 6 ECHR: Fair Trial
  The concept of fair trial refers to issues like presumption of innocence, the right to be informed promptly of the cause and nature of an accusation, and the right to defend oneself in person.
  The use of algorithms in crime prevention or anti-terrorism may potentially limit fair trial for some individuals, especially when they are used to predict their future behaviour.
- Art. 8 ECHR: Right to Privacy
  The right to private life may be increasingly threatened by the capability of algorithms to handle large volumes of data of various types (text, images, videos…).
  The EU General Data Protection Regulation (GDPR) framework that will enter into force in May 2018 aims to better protect the individuals against misuse of the data related to them.
- Art. 10: Freedom of Expression
  Online media usually use algorithms to predict users' preferences and to filter content accordingly. This selection may limit the awareness of diverging opinions and the promotion of diversity. Furthermore, content removal policies may filter information on the basis of questionable principles. For instance, removing "extremist" videos should rely on a clear definition of "extremism".

- Art. 11: Freedom of Assembly and Association
  Some algorithms may be used to classify and sort out some people from calls for assemblies, popular demonstrations or peaceful protests, which would limit their rights.
- Art. 13 Right to an Effective Remedy
  This right refers to the right to a reasoned and individual decision in a complain procedure. Companies are increasingly using algorithms to handle complaint procedures. Only a fraction of cases is selected to be handled by a human operator. Biases in the results of the algorithms used to identify and select those cases may infringe the right to an effective remedy.
- Art. 14 Prohibition of Discrimination
  Search algorithms are selecting information and are providing a ranked list of results according to their perceived relevance for the users (filter bubble effect). This may cause discrimination among people, especially in case of quasi-monopolistic position of a search engine (like Google).
- Social Rights and Access to Public Services
  If algorithms are used to manage people in an organisation (e.g. recruitment, evaluation, dismissal), new concerns emerge concerning the lack of transparency of these automated decisions. It also raises the risk of abusively considering the decisions to be purely objectives. Similar concern can be pointed out in the case of algorithmic decisions made by public services. Discrimination regarding the access to social services could result from biased algorithms.
- Right to Free Elections
  At the time of elections, filter bubbles on social media and search engines can be used to influence the democratic process. Providing only the exact content that a citizen is expecting is an easy way to reinforce his opinions. Let's note here that some companies like Cambridge Analytica [66] are publicly selling data-driven algorithmic services to support election campaigns. Very recently Facebook policy regarding political ads [67] has also been questioned.

## Initiatives from companies

There is a growing awareness within private enterprises to include ethics in their initiatives involving decision making algorithms. It may take various forms, from internal policy to setting up alliances, consortiums or partnerships. Some examples are listed in this section.

### Partnership on AI

In 2016, 7 major Anglo-Saxon companies offering product and services based on AI (Amazon, Apple, DeepMind, Google, Facebook, IBM, Microsoft) gathered and founded the "*Partnership on AI to benefit people and society*" [68]. This initiative pursues four goals:

- develop and share best practices;
- advance public understanding;
- provide an open and inclusive platform for discussion and engagement;
- identify and foster aspirational efforts in AI for socially beneficial purposes.

Since 2016, several companies (e.g. Salesforce.com, Sony, SAP, Zalando…) and organisations (e.g. Human Rights Watch, UNICEF, Electronic Freedom Foundation, Amnesty International…) have joined this partnership.

### IBM

The official IBM Policies and Principles available online [69] include a section entitled "*IBM Cognitive Principles*" related to the ethical guidelines with regard to the use of Artificial Intelligence (which includes ML). Three major topics are addressed:

- purpose: designing systems that augment human intelligence, but which ultimately remain under human control;
- transparency: making clear when and for what purpose AI is used, providing information about the data and the methods used, working with clients to protect their data and insights;
- skills: helping the students, workers and citizens to develop the skills and knowledge to benefit from the cognitive economy.

### DeepMind

The DeepMind British company has been acquired by Google in 2014. It has recently created a unit called "*DeepMind Ethics & Society*" because "*technology is not value neutral, and technologists must take responsibility for the ethical and social impact of their work*" [70]. They have published online the five principles governing their work on AI:

- social benefit;
- rigorous and evidence-based;
- transparent and open;
- diverse and interdisciplinary;
- collaborative and inclusive.

### Audi

Audi car manufacturer has recently launched the so-called "*Beyond Initiative*" [78] aiming to set up an interdisciplinary network of AI experts. The rationale of this initiative is that business actors, politicians and civil society must work together to shape the transformations brought by AI technologies. In his speech during the "*AI for Good Global Summit*" held in Geneva in June 2017, Prof. Stadler, CEO of Audi has explained that the "*Beyond Initiative*" aims to tackle the ethical and legal questions of piloted and autonomous driving (where ML plays a major role). He also highlighted the impact of AI on the working world: mechanisms of the labour market, collaboration between humans and machines.

## Initiatives from other stakeholders

Some professional associations or other types of similar organisations have also identified the need to encompass the ethical facets of applications using AI and ML. Some examples are provided below.

### IEEE

IEEE (Institute of Electrical and Electronics Engineers) is one of the largest and most famous professional associations for computer science and engineering. In 2016, IEEE has launched an initiative on Ethical Considerations in Artificial Intelligence and Autonomous Systems [74]. Its official mission is "*to ensure every technologist is educated, trained, and empowered to prioritize ethical considerations in the design and development of autonomous and intelligent systems.*" In a document entitled "*Ethically Aligned Design: A Vision for Prioritizing Human Wellbeing with Artificial Intelligence and Autonomous Systems (AI/AS)*" three principles are stated:

- *Embody the highest ideals of human rights that honour their inherent dignity and worth;*
- *Prioritize the maximum benefit to humanity and the natural environment;*
- *Mitigate risks and negative impacts as AI/AS evolve as socio-technical systems*.

The rationale of this initiative is to foster the development of norms and standards about ethical governance of AI/AS.

## ITU

In June 2017, the International Telecommunications Union (ITU) and the XPrize Foundation (supported by IBM) have organized the "*AI for Good Global Summit*" in collaboration with several Agencies of the United Nations (e.g. UN Human Rights Office of the High Commissioner, UNICEF, UN Framework Convention for Climate Change) [76]. The general purpose of the summit was to address how AI can help to achieve the UN Sustainable Development Goals (SDG). Indeed, AI can underpin solutions that tackle challenges related for instance to poverty, hunger, health, or the preservation of the environment. More than 500 speakers from science, business, governments and non-profit organizations expressed their views during the summit. Although the question of standards and regulation was debated, an observer of the Summit [77] reported that "*Overall, the Summit made clear that it is too early to formulate definitive guiding principles for AI that have the highest chance to make the world a better place. The event generated a lot of healthy food for thought, which will be followed up on by smaller working groups whose task will be to digest the discussions and turn them into the first U.N.-backed AI guiding principles*."

## FAT-ML

The Fairness, Accountability and Transparency in Machine Learning (FAT-ML) community [72] is proposing a series of principles for algorithmic accountability and social impact considerations. Since 2014, it organises a scientific interdisciplinary workshop on this topic. The FAT-ML web site publishes some "*Principles for Accountable Algorithms and a Social Impact Statement for Algorithms*" [73]:

- responsibility,
- explainability,
- accuracy,
- auditability,
- fairness.

Compared to other similar list of theoretical principles, FAT-ML also provides some guiding questions helping to take concrete actions in an operational context. For instance, regarding the "*accuracy*" principle, they suggest answering the following questions:

- "*What sources of error do you have and how will you mitigate their effect?*
- *How confident are the decisions output by your algorithmic system?*
- *What are realistic worst case scenarios in terms of how errors might impact society, individuals, and stakeholders?*

- *Have you evaluated the provenance and veracity of data and considered alternative data sources?"*

The list of questions proposed by FAT-ML may be a good entry point for companies that want to assess their current position regarding decision-making algorithms.

## Royal Society

The British Royal Society is among the most prestigious learned society for science and possibly also the oldest (founded in 1660). In 2017, it has published a report on "*Machine learning: the power and promise of computers that learn by example*" [75] in which some concerns about ethical issues are mentioned. Most of the previously issues are also identified by this report, such as privacy, fairness, interpretability, or transparency.

Interestingly, this report points out the need to include the ethical and social implications of ML within teaching activities, not only for computer scientists but also for students enrolled in social or health related programmes. More broadly speaking, the Royal Society advocates for the introduction of ML lessons sooner in the education of the children. Ethical questions as well as the issue of false negative and false positive is the last topic address in the proposed educational path. The final purpose would be to provide the new generation of citizens and workers with the sufficient skills to really benefit from ML widespread use and not to suffer from this evolution.

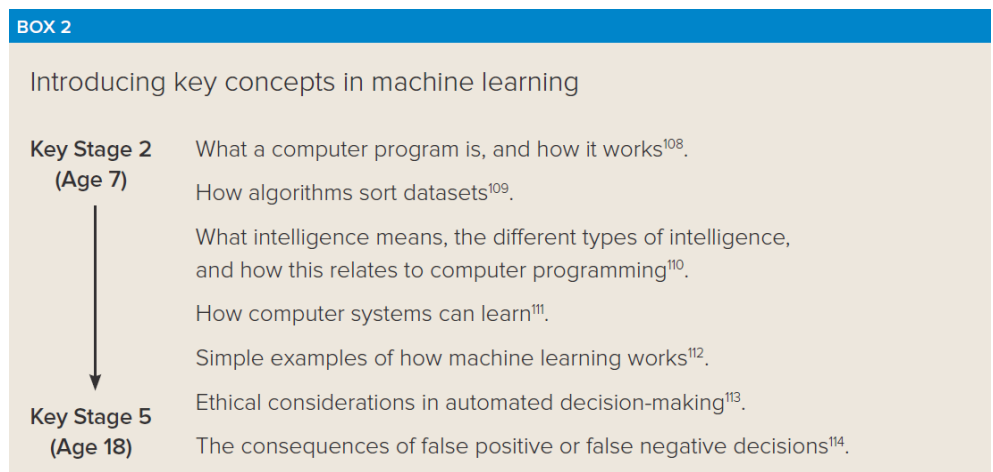

Figure 17: Step-by-step introduction of ML concepts in education (source: [75])

The Royal Society report also speaks for the creation of governance systems to address social and ethical challenges in this domain.

## Key aspects of ongoing initiatives

We have illustrated in the previous sections that the ethical questions raised by decision-making algorithms are progressively receiving greater attention from numerous stakeholders who have identified the need to launch or to join some initiatives.

The public authorities need to better understand if this new technology will lead to a socio-economic paradigm shift. Some public bodies have therefore initiated some initiatives to better assess how decision-making algorithms will influence their scope of activities. The Council of Europe is very much concerned by Human Rights and it has tried to map the risks of ML accordingly. The European Union or the UK Parliament need to better assess the global influence of ML on the society. In the USA, the DARPA agency is dealing with military and security related issues. Consequently, DARPA aims to see the potential and the limitations of decision-making algorithms from this perspective.

Private companies can be distinguished according to their current or future position towards decision-making algorithms. Some companies like Google, Facebook or IBM are actively working for several years on the development of ML-based systems. They have acquired valuable technological knowledge about it as well as how to derive business value out of it. They are also aware of the weaknesses of such systems. They are thus doing research to try to limit the annoying behaviours and outcomes of decision-making algorithms that could ultimately lead to restriction imposed on their usage.

Other companies like Audi have identified the disruptive potential of decision-making algorithms on their business. Although they are not pure players of the ICT sector, they have started to integrate or at least to explore how to integrate ML algorithms in their product and services. They have also realized that such an evolution does not only require to solve technological challenges. Ethical issues will also play a major role regarding both what they are allowed to bring to the market and what the customers are willing to accept.

Finally, professional associations, consortia, and scientific societies are also taking steps towards providing support to their members regarding how to deal with the ethical questions raised by a wider diffusion of decision-making algorithms in our society.

We could say that the organisations (public and private) seem to follow a step-by-step process before adding ethics on their agenda regarding decision-making algorithms. The very first step is to be aware of the existence of decision-making algorithms. Next the organisations are assessing if, how and when this technological evolution can be somehow beneficial or rather threatening for them. When the technological challenges and the business opportunities are better understood, they start thinking about the conditions under which decision-making algorithms can be used in commercial products or public services. Among these conditions, the regulatory framework is often studied first. Finally, they are progressively convinced that legal compliance is not sufficient and ethics appears as a new challenge to be seriously managed.

# Regulation

## Current situation

We have explained in a previous section that various public bodies and authorities have launched some initiatives to better understand the very nature of decision-making algorithms. They also want to raise awareness about the risks this disruptive technology may generate for the society. Beside these initiatives, the public authorities are also in charge of setting up the appropriate regulatory framework. Regulating obviously means going further than simply understanding or raising awareness. This explains why we are discussing it in a specific section.

In their second draft version of the report on the Human Rights Dimensions of Algorithms [23], some experts selected by the Council of Europe explain that "*there is no normative framework for the development of systems and processes that lead to algorithmic decision-making or for the implementation thereof. In fact, it is unclear whether a normative framework regarding the use of algorithms or an effective regulation of automated data processing techniques is feasible as many technologies based on algorithms are still in their infancy. Issues arising from the use of algorithms as part of the decision-making process are manifold and complex and include concerns about data quality, privacy and unfair discrimination. At the same time, the debate about algorithms and their possible consequences for individuals, groups and societies is at an early stage.*"

However, many technologies relying on decision-making systems are in operational use today and are somehow regulated. For instance, autopilot systems in aircraft has long had to be approved by regulators. Similarly, trading algorithms are regulated by the MiFID II directive. In many countries like in Australia and New Zealand, the software embedded in 'slot machines' are regulated by the government and must be fair, secure and auditable.

The key question is then to know if we will face an increase in regulation frameworks to handle all applications fields of algorithmic decision-making. Latzer et al. [9] explain that a large range of options are possible to limit the risks and to increase the potential benefits, from market mechanisms to control regulation by public authorities. In between "*there are several additional governance options: self-organization by individual companies; (collective) industry self-regulation; and co-regulation – regulatory cooperation between state authorities and the industry*" [9].

The governance of decision-making algorithms is not limited to investigations about the algorithm behaviour itself. Organisational settings and the interplay between various stakeholders is also part of the governance. For instance, how the data feeding a machine learning algorithm are collected, cleaned and checked is as important as which ML technique is used to derive a decision. These aspects should also be considered in a regulatory framework of decision-making algorithms.

The emergence of a (ground-breaking) technology like ML does not require automatically to set up new regulation schemes. Most of the existing regulation frameworks are still relevant to deal with some facets of decision-making algorithms. For instance, competition law set rules to specify how market players should behave with each other as well as how they can interact with customers. Cartels, abusive concentration, illegitimate entry barriers can also be encountered in the stakeholders of the decision-makers algorithms. However, a new technology may also lead to new types of issues that were never met before. We have explained in the previous chapters that decision-making algorithms are generating new fundamental questions, especially in terms of ethics.

In 2018, the EU General Data Protection Regulation (GDPR) will change the rules of the game for the stakeholders collecting, storing and processing data [65]. Data subjects will have the right not to be the object of decision made solely on the basis of automated individual decision-making. Some exceptions are admitted but all stakeholders will have to carefully check if they apply in their respective case. The data controller will have the obligation of fairness and transparency when it processes data. In terms of information to the data subject, giving meaningful information about the logic involved in the processing will make many professionals' hair turn grey if "black-box" algorithms are used to take decision.

It is important in this debate to keep in mind how realistic is the regulatory framework and if it can be implemented in practice. For instance, Mulgan [3] explains that *we shouldn't expect too much from transparency, which can be hard to make the most of without what are still scarce technical skills*."

Regarding the impact of GDPR in practice, Mittelstadt at al. [25] also adopt a more critical point of view. According to these authors, the wording of the regulation allows it to be either "*a toothless or a powerful mechanism to protect data subjects dependent upon its eventual legal interpretation*". At the end of the day, the supervisory authorities and their decisions will determine GDPR effectiveness. It seems clear that additional work is still needed to provide normative guidelines and practical mechanisms to help the stakeholders to put this new regulation into practice.

## Suggested evolutions in regulation

Considering the huge impact that decision making algorithms will increasingly have on people's life and on business activities, the debate about its regulation is very active and will for sure still grow in the coming years. For illustrative purposes we discuss below two proposals chosen among many others. We have selected them because the Anglo-Saxon mindset is usually less favourable to regulation of any kind than the approach promoted in Continental Europe. It is therefore even more striking to observe Anglo-Saxon authors speaking for setting up public bodies to regulate the field.

### NESTA proposal

G. Mulgan, from the NESTA UK-based foundation, suggests the creation of a so-called "Machine Intelligence Commission", defined as "*a new public institution to help the development of new generations of algorithms, machine learning tools and uses of big data, ensuring that the public interest is protected*" [3]. According to Mulgan, this public body is a prerequisite to increase informed trust from the public. His approach is not only founded on moral principles, but he also argues that trust in technologies is key to fully benefit from them from a socio-economic perspective. He reminds that new technologies usually bring new ethical and moral issues that progressively give birth to new legal or regulatory frameworks. It has been the case for nuclear energy, television or genetically modified organisms. There is no reason for decision-making algorithms to be differently treated. Although setting strict rules of the game may appear counter-productive for the early developers of a technology, it is in fact a condition to build a sustainable market around it in the longer term. None of the extreme behaviours (overoptimistic vs. paranoid) is appropriate. Mulgan proposes the Machine Intelligence Commission to have "*strong powers of investigation and of recommendation*" but no "*formal regulatory powers of approval or certification*". He admits that its role will very likely evolve with time. This Commission would need

*"strong legal, social science and design capabilities, as well as technical capabilities in data, information architectures and business models."*

## An FDA for Algorithms

Andrew Tutt [6] argues for the creation of a new US agency focused on Algorithms. As a lawyer his main motivation is on "*how best to prevent, deter, and compensate for the harms that they [the algorithms] cause*". He wants to prevent unacceptably dangerous algorithms to go to the market. Prior approval by this so-called "FDA for Algorithms" would be required for some types of algorithms to be distributed or sold.

Tutt focuses on machine-learning algorithms that will "*pose significant risks to individuals and society if they fail or are misused*" [6]. As examples he mentions such algorithms used in self-driving cars, medical diagnoses, or power grids. A key concern for him is the fact that it is very difficult to predict when and how they can fail. In his opinion, "*how machine-learning algorithms learn—and how they reason from experience to practice—is almost entirely alien*". This statement is only partly right because the degree of predictability of the results depends on the specific algorithm that is considered. Nevertheless, it perfectly illustrates how difficult it may be for non-experts in machine-learning to precisely know what is understandable and explainable in the reasoning process.

Tutt identifies two major issues with ML algorithms: predictability and explainability [6]. "*Algorithm's predictability is a measure of how difficult its outputs are to predict, while its explainability is a measure of how difficult its outputs are to explain.*" [6] We have shown in previous chapters that this vision is widely acknowledged in the community. Tutt also argues that ML algorithms raise critical (still unsolved) questions in terms of responsibility. How to measure that an algorithm has acted negligently or in a legally culpable way? How to trace that an algorithm has behaved according to legal standard? How to distribute the responsibility among the humans who have played a role in the process bringing a ML algorithm to operational use?

The FDA for Algorithms proposed by Tutt would have several roles:

- Acting as a standard-setting body:
    - classifying algorithms according to a standard scale,
    - establishing standard methods to test the algorithm performance,
    - developing design standards,
    - developing standards for distributing liability among coders, implementers, distributors, and end-users.
- Acting as a Soft-Touch Regulator:
    - imposing requirements of openness, disclosure, and transparency.
- Acting as an Hard-Edged Regulator:
    - pre-market approval.

# Analysis of Corporate Reports

We have explained in the previous chapter that ML algorithms have a great potential to solve complex problems but in the same time they must be used with care by informed persons. Several risks are associated to inappropriate applications of ML-based systems.

Machine Learning is already deployed in production in several companies and organizations. Considering the risks that it may cause on their stakeholders (employees, customers, suppliers, the society, the environment...) we could consider that they should report on it. How are they using ML? Why do they use ML? Do they have established some rules or guidelines regarding ML use? Are ML systems used by used informed about the associated risks? Are the ML systems designed in-house or purchased to a supplier? Numerous questions can be relevant to assess the risk profile of an entity due to the use of ML systems? In a previous section (cf. p. 18) we have demonstrated that even the most advanced IT companies or the most powerful governmental agencies have faced issues caused by a ML system that went out of control or delivered annoying outcomes.

In this chapter, we have examined the annual reports of four companies. Because they are among the technological leaders in the field, we have selected Google (Alphabet) and IBM. We have also studied two other companies, namely Salesforce and Zalando because that are publicly taking initiatives to foster the integration of ML-based applications in their products and services.

To provide comparable results, we have analyzed the reports with the same methodology.

- Download from the web the annual report (in English) for the year 2016.
- Search for the expressions (*machine learning*, *artificial intelligence, analytics, predict / prediction / predictive / predictable*) in the report with the full text search function of Acrobat Reader. Titles, headers and footers are excluded from the search results.
- Read the sections of the annual report where the searched expressions where found.
- Check if some risks related to ML or AI are mentioned in the sections identified at the preceding step.
- Search for the expressions (*risk*, *risks*) in the report with the full text search function of Acrobat Reader. Titles, headers and footers are excluded from the search results.
- Check if the sections where "risk" or "risks" are mentioned are referring to AI or ML issues.
- Summarize and comment the results.

A more compete study could have analyzed the reports more thoroughly with more advanced text analytics techniques but by lack of time it has not been possible in the context of this paper. One might also argue that the annual report of a company is not the right place to discuss the risks related to ML that we have identified in this paper. For sure, other documents should also be considered to get a better picture of a (large) company. Our study must be seen as a first step of a larger process that would require much more time and resources. The initial results presented hereafter should therefore be confirmed by future research.

## Alphabet (Google)

Alphabet, Inc. clearly states that it is investing and using AI and ML: "*Across the company, machine learning and artificial intelligence (AI) are increasingly driving many of our latest innovations. Within Google, our investments in machine learning over a decade are what have enabled us to build Google products that get better over time, making them smarter and more useful…*" [56] Larry Page, CEO of Alphabet, explains in the 2016 annual report [56]: "*Sundar [Pichai] is doing great as Google CEO. […] But, I'm excited about how he is leading the company with a focus on machine learning and AI*".

Indeed, Google stands at the forefront of research in AI and ML. In fact, the company originates from the page rank algorithm that belongs to this domain of computer science. Numerous papers have been published by its researchers in this domain. Of course, these scientific papers are read by a (very) small fraction of the population. Beside this community of experts, how does Google report on its use of ML in its business? We have explored the 2016 Annual Report of Alphabet, Inc [56] (the listed mother company of Google, created in 2015) to have an initial idea about it.

| **Expression** | **Frequency (# of occurrences)** | **Reference to associated risks** |
|---|---|---|
| *machine learning* | 9 | None, only positive statements |
| *artificial intelligence* | 1 | None, only positive statements |
| *analytics* | 1 | None, only positive statements |
| *predict / prediction / predictive / predictable* | 7 | No reference to ML/AI related risks<br>References to<br>- business and market risks<br>- financial statements<br>- claims and investigations |

| **Expression** | **Frequency (# of occurrences)** | **Reference to AI / ML issues** |
|---|---|---|
| *Risk(s) / Risky* | 56 | Reference to development of risky activities, projects, products or services<br><br>*Indirect references to AI / ML:*<br>Risks due to new regulations<br>Risk of product liability<br>Risk of intellectual property litigation<br>Risk related to data privacy<br>Risk of cyber-attacks<br>Risk linked to trademarks<br>Risk of quality issues in the software products |

Table 3: Analysis of 2016 Alphabet Annual Report

On the basis of this initial analysis, we can draw some preliminary yet interesting conclusions. First, although Alphabet is officially placing Machine Learning at the heart of its products and services, the potential risks associated to the usage of this technology are not explicitly mentioned in its annual report. To be fair, we must admit some indirect references but very similar to what any software company would declare, should it use ML or not.

© Benoît Otjacques, BECM programme 2016-2017, LSM    51

## IBM

The first page of the 2016 Annual Report of IBM [57] explains that the company is reinventing itself *"to fuel your dreams with Watson, with IBM Cloud, with deep expertise, with trust"*. Without going into technical details, Watson is the name of IBM software suite offering analytics services (notably based on AI and ML). According to some observers, AI is at the core of IBM strategy. For instance, Jason Bloomberg explains in Forbes that *"IBM Bets The Company On Cloud, AI And Blockchain"* [84]. Therefore, it is meaningful to further explore how IBM is communicating on the risks associated to AI and ML in its annual report.

| Expression | Frequency (# of occurrences) | Reference to associated risks |
|---|---|---|
| *machine learning* | 3 | None, only positive statements |
| *artificial intelligence* | 6 | None, only positive statements |
| *analytics* | 53 | Almost only positive statements<br>Reference to financial risks (business development of IBM analytics products and services is not compensating yet the decline of traditional business lines)<br>IBM Analytics can be used to help clients assess their risk and compliance against industry guidelines<br>References to IT security risks |
| *predict / prediction / predictive / predictable* | 18 | References to<br>- business and market risks<br>- financial statements<br>- claims and investigations<br>- positive statements about use cases in health, industrial maintenance, agriculture, insurances, police |

| Expression | Frequency (# of occurrences) | Reference to AI / ML issues |
|---|---|---|
| *Risk(s) / Risky* | 100+ | References to cases where AI/ML can reduce risks<br><br>*Indirect references to AI / ML:*<br>Risk related to data privacy<br>Risk of cyber-attacks |

Table 4: Analysis of 2016 IBM Annual Report

To summarize, in its 2016 annual report IBM does not draw the reader's attention to the risks associated to increasing use of decision-making algorithms. On the contrary, IBM explains several times how the use of its analytics solutions can lower some business risks (and it is very often true). However, since IBM is a leading company in the AI / ML field for many years, it is very unlikely that none of its brilliant experts be aware of biases and risks of potentially unfair decisions taken by such algorithms. For instance, the example of analytics for proactive crime prevention ([57] p. 30) does not make any reference to the related ethical issues. Similarly, some statements may be misleading for readers who are not experts in computer science: *"Cognitive systems are not programmed; like humans, they learn from experts and from every interaction, and they are uniquely able to find* *patterns in big data. They learn by using advanced algorithms to sense, predict and infer.*" Such a statement may give the impression that ML systems are very much like humans, but it hides their intrinsic limitations which are of different nature of the humans' ones. To be fair, we must remind here that IBM has published an official policy (cf. p. 43) including ethical aspects. This is however a good example to illustrate how beneficial an integrated report (including ML-related ethics and risks) can be to ease the complete assessment of a company.

## Zalando

Zalando is an online fashion retailer that is leader on the European market. At the Internet Retailing Summit held in Berlin in June 2017, Andreas Antrup, Vice President, data and advertising at Zalando gave an interview about innovation in his company [85] in which he identifies machine learning and artificial intelligence as key factors to drive innovation. In 2016, Zalando has launched the Muze experimental project together with Google to explore the potential of ML in the fashion industry [87]. The idea was to ask a ML-based system to design individual clothes based on the personal preferences of the customers. Zalando has also set up an internal research lab "*to uncover advancements in data science, machine learning, and artificial intelligence and to explore their applications to fashion commerce*"[86]. It seems therefore that ML is on Zalando's agenda. Although it is not a pure IT company, it may nevertheless be instructive to see if it has started to report on its use of ML and the related risks.

| **Expression** | **Frequency (# of occurrences)** | **Reference to associated risks** |
|---|---|---|
| *machine learning* | 1 | None, only a statement about the creation of a research lab on ML and AI |
| *artificial intelligence* | 1 | None, only a statement about the creation of a research lab on ML and AI |
| *analytics* | 2 | None |
| *predict / prediction / predictive / predictable* | 5 | References to<br>- business and market risks<br>- financial statements |

| **Expression** | **Frequency (# of occurrences)** | **Reference to AI / ML issues** |
|---|---|---|
| *Risk(s) / Risky* | 200+ | Detailed description of Governance, Risk & Compliance management (not specific to AI/ML)<br><br>*Indirect references to AI / ML:*<br>Risks of unavailability of IT systems<br>Risks due to new regulations |

Table 5: Analysis of 2016 Zalando Annual Report

The 2016 annual report of Zalando provides a very detailed chapter about risk management in the company. It also include a specific GRI (Global Reporting Initiative) section in which are listed several topics where decision-making algorithm may play a role, like Non-Discrimination, Human Rights, or Privacy. At this stage, the absence of references to ML-related risks and issues is not shocking because it seems that Zalando is just at the beginning of a large use of decision-making algorithms that may potentially raise ethical issues. We remind here that the optimization of plant operations or logistics is out of the scope of this paper. In the next years it would be interesting to analyze the annual reports of Zalando to see if the risks associated to decision-making algorithms will appear, especially in the GRI section.

## Salesforce

Salesforce is a market leader in software applications for Customer Relationship Management (CRM). In the introduction of its 2016 Annual Report [59], Salesforce provide a vision for its future: "*And we're moving into the future with the same focus, bringing you **data science, analytics**, and the Internet of Things to name just a few*." A clear reference to ML is also included later in this document: "*Progress in **data science** and **machine learning** is moving companies beyond just automating business processes to more data-driven, predictive computing solutions.*" In June 2017, Salesforce has launched Einstein Analytics that adds AI-based features to the Salesforces CRM (Customer Relationship Management) system. It is therefore relevant to wonder how Salesforce reports on AI / ML and how the company reports on possible associated risks.

| Expression | Frequency (# of occurrences) | Reference to associated risks |
|---|---|---|
| *machine learning* | 2 | None<br>General statement & Financial Statement related to the acquisition of stocks of RelateIQ, Inc |
| *artificial intelligence* | 0 | |
| *analytics* | 14 | Almost only positive statements<br>Reference to financial risks (business development of analytics products and services may not be successful) |
| *predict / prediction / predictive / predictable* | 10 | References to<br>- business and market risks<br>- financial statements<br>- new data protection laws and regulations |

| Expression | Frequency (# of occurrences) | Reference to AI / ML issues |
|---|---|---|
| *Risk(s) / Risky* | 48 | *Indirect references to AI / ML:*<br>Risk of product liability<br>Risk of intellectual property litigation<br>Risk related to data privacy<br>Risk of cyber-attacks<br>Risk of quality issues in the software products |

Table 6: Analysis of Salesforce 2016 Annual Report

We must admit that Salesforce is probably less advanced than Alphabet (Google) or Amazon in the integration of ML in its commercial products. Although the company has started to take initiatives in this domain in the last years, it is understandable that it does not intensively report on it (yet). It will be very interesting to observe how ML-related products and services will evolve within Salesforce. Indeed, the nature of the data handled by a CRM system can potentially lead to a slippery slope towards ethical issues.

# Conclusions

> *"It is thus too simple to blame the algorithm or to suggest to no longer resort to computers or computing. Rather, it is the social construct and the specific norms and values embedded in algorithms that need to be questioned, criticised and challenged.*
> *Indeed, it is not the algorithms themselves but the decision-making processes around algorithms that must be scrutinised in terms of how they affect human rights."* [23]

Some decades ago, Nobel prize- as well as Turing[1] prize-winner Herbert Simon has theorized the concept of bounded rationality of human beings [88]. Indeed, people are often biased when they take decisions and they sometimes behave in an irrational manner. Is this observation constituting a solid enough argument to replace humans to take decisions wherever technology allows it? This fundamental question has motivated the work described in this paper. Nowadays, replacing humans by algorithms to take business decisions is frequently presented as the new Holy Grail but are these algorithms as unbiased, fair and trustworthy as they are supposed to be?

Of course, we acknowledge that the impressive recent development of machine learning algorithms will in many cases be highly valuable. ML-based systems help to save lives by providing faster diagnosis. They help companies to fight frauds and abuses. Environmental disasters are better predicted thanks to ML-based analysis of satellite images. We are undeniably living at the beginning of a new era where humans will delegate some decisions to algorithms at a pace never encountered before.

We believe, however, that such a ground-breaking technology is not by definition safe. In the past, after a pioneer age where exotic experiments were carried out and basic rules were discovered, most of the game changing technologies have been standardized and regulated. On one hand, it has given some safeguards for the citizens and the consumers. On the other hand, it has clarified the rules of the game for the companies which has given them more visibility to draw their roadmap. Such an evolution has been observed when new forms of energy (e.g. electricity) or new transportation means (e.g. automobile) have demonstrated their advantages on former technologies. Although it is still unclear how decision-making algorithms could be and should be regulated, it is vey likely that they will increasingly be in the coming years.

Many myths are propagated regarding artificial intelligence and machine learning. They are underpinned sometimes by ignorance, sometimes by misconduct. Indeed, the underlying concepts and the related technologies are not trivial to understand for non-experts. Nevertheless, this inherent complexity can not justify misleading statements, deliberate underestimation of risks or unfair decisions.

We believe that critical thinking will become even more required than it has been in the past. Statements like "*only what can be measured can be managed*" or *"according to our prediction this person will behave like this"* should be scrutinized according to how data have been collected, pre-processed and used to build a predictive model. Too often the notions of correlation and causation are confused with each other. Many people have difficulties to grasp the real meaning of a probabilistic result. We must also keep in mind that machine learning algorithms are not exact replications of the human brain. They are different in nature even if they can provide acceptable solutions to similar problems.

---
[1] ACM Turing prize is considered as the Nobel prize of computer science.

Companies are increasingly asked to report on the defense of human rights, the protection of the environment or the fight against crime. We believe that corporate reports will also include references to risks related to decision-making algorithms in the future. Unfortunately, reporting on the use of decision-making algorithms for a public of non-experts is very difficult. The analysis of few reports of companies actively involved in this evolution has shown that they are very prudent to associate artificial intelligence or machine learning to (ethical) risks. We don't have any doubts that leading companies like Google or IBM are fully informed about these risks, but we fear that numerous other enterprises that are about to use decision-making algorithms are simply not aware of the related risks. Let's keep in mind in this context that machine learning systems are ideal candidates to trigger Taleb's "black swans" (rare and unpredictable events having extreme impact).

Finally, this paper is only a very preliminary work on the ethical issues raised by decision-making algorithms and how to report on them. We have really enjoyed exploring this emerging topic. Combining recent advances in technology, legitimate business needs and indispensable ethical thinking is a fascinating undertaking.

At the end of this paper, we leave the reader with some food for thought.

> Will we have to deal with "bloody data" in the future
> just like we have to with "conflict minerals" today?

## Acronyms

List of acronyms used in this paper.

- ACM: Association for Computing Machinery
- AI: Artificial Intelligence
- DMA: Distributed Moral Action
- DMR: Distributed Moral Responsibility
- ECHR: European Convention on Human Rights
- GDPR: General Data Protection Regulation
- GRI: Global Reporting Initiative
- HRDAG: Human Rights Data Analysis Group
- ICRAC: International Committee for Robot Arms Control
- ITU: International Telecommunication Union
- IEEE: Institute of Electrical and Electronics Engineers
- ML: Machine Learning
- NGO: Non-Governmental Organization
- NSA: National Security Agency
- UN: United Nations
- UNSDG: United Nations Sustainable Development Goals